\documentclass[12pt,reqno]{article}
\usepackage{amsmath}
\usepackage{amsthm}
\usepackage{amsfonts}
\usepackage{amssymb}
\usepackage{epsfig}
\usepackage{graphics}
\usepackage{graphicx,indentfirst}
\usepackage{hyperref}
\numberwithin{equation}{section}

\def \Fig#1#2#3 {
\begin{figure}
\centering
\epsfxsize=#2cm \epsfbox{#1.eps}
\caption{#3}
\label{#1}
\end{figure}
}

\newcount\figno
\def\fig#1#2#3{
\par\begingroup\parindent=0pt\leftskip=1cm\rightskip=1cm\parindent=0pt
\baselineskip=15pt
\global\advance\figno by 1
\epsfxsize=#3
\centerline{\epsfbox{#2}}
\vskip 12pt
{\bf \small Figure \the\figno:} {\small #1}\par
\endgroup\par
}
\def\figlabel#1{\xdef#1{\the\figno
\mbox{ }}}
\def\encadremath#1{\vbox{\hrule\hbox{\vrule\kern8pt\vbox{\kern8pt
\hbox{$\displaystyle #1$}\kern8pt}
\kern8pt\vrule}\hrule}}

\def\L2{{\it Fun\/}\bigl(\text{\GL}\bigr)}

\newcommand{\Om}{\Omega}
\newcommand{\bm}{b_-}
\newcommand{\bp}{b_+}
\newcommand{\cm}{c_-}
\newcommand{\cp}{c_+}

\newcommand{\del}{\partial}

\newcommand{\ep}{\epsilon}

\newcommand{\ad}{\text{ad}}

\def\agl{{\rm $\widehat{\text{gl}}$(1$|$1)}}
\def\gl{{\rm gl(1$|$1)}}
\def\GL{{\rm GL(1$|$1)}}
\newcommand{\rrangle}{{\rangle\!\rangle}}
\newcommand{\llangle}{{\langle\!\langle}}

\newcommand{\Z}{\mathbb{Z}}

\newcommand{\str}{\text{str}}

\newcommand{\tr}{\text{tr}}
\newcommand{\<}{\langle}
\renewcommand{\>}{\rangle}

\def\agl{{\rm $\widehat{\text{gl}}$(1$|$1)}}
\def\gl{{\rm gl(1$|$1)}}
\def\GL{{\rm GL(1$|$1)}}

\providecommand{\abs}[1]{\left\lvert#1\right\rvert}

\newcommand{\al}{\alpha} \newcommand{\ga}{\gamma}
\newcommand{\Ga}{\Gamma} 
 \newcommand{\de}{\delta}
 \newcommand{\si}{\sigma}
\newcommand{\la}{\lambda} 

 \newcommand{\De}{\Delta}

\newcommand{\half}{\mbox{$\frac{1}{2}$}}

\providecommand{\abs}[1]{\left\lvert#1\right\rvert}

\newcommand{\nn}{\nonumber}

\newcommand{\expect}[1]{\langle #1\rangle}

\newcommand{\normord}[1]{{} : #1 : {}}

\newcommand{\delbar}{\bar{\partial}}
\newcommand{\zbar}{\bar{z}}

\author{Thomas Creutzig and Peter B. R\o nne\\[5mm]
    DESY Theory Group, DESY Hamburg  \\
Notkestrasse 85, D-22607 Hamburg, Germany \\
\phantom{wwwx}{\small e-mail: }{\small\tt
thomas.creutzig@desy.de, peter.roenne@desy.de} }

\date{December 2008}

\begin{document}
\begin{titlepage}
    \title{\Huge The \GL-symplectic fermion correspondence}
    \maketitle       \thispagestyle{empty}

\vskip1cm
\begin{abstract}
In this note we prove a correspondence between the
Wess-Zumino-Novikov-Witten model of the Lie supergroup \GL\ and a free
model consisting of two scalars and a pair of symplectic fermions. This
model was discussed earlier by LeClair. Vertex operators for the
symplectic fermions include twist fields, and correlation functions of
\GL\ agree with the known results for the scalars and symplectic fermions.
We perform a detailed study of boundary states for symplectic fermions and
apply them to branes in \GL. This allows us to compute new amplitudes of
strings stretching between branes of different types and confirming
Cardy's condition.
\end{abstract}
\vspace*{-16.9cm}\noindent {\tt {DESY 08-195 }}

\end{titlepage}

\section{Introduction}

Conformal field theories with supersymmetric target space has become an
important area of current research. They are essential in a variety of
significant problems both in string theory and in disordered systems.

Understanding sigma models on supersymmetric spaces deep in the strongly
coupled regime is of primary importance. In many models one believes that
there exists a dual description which is better accessible in such a
regime. The most prominent example is certainly the celebrated AdS/CFT
correspondence \cite{Maldacena:1997re,Aharony:1999ti}, but there are, of
course, other interesting dualities involving sigma models on
supersymmetric spaces. For example, recently a strong-weak duality between
the OSp(2N+2$|$2N) Gross-Neveu model and the principal chiral model on the
supersphere $S^{2N|2N+1}$ was conjectured
\cite{Candu:2008yw,Candu:2008vw}.

There are various ways to find and to test such correspondences. Many
supersymmetric spaces possess a family of conformally invariant field
theories, and points in the moduli space that are exactly solvable e.g.
the Wess-Zumino-Novikov-Witten point on supergroup manifolds or the
infinite radius limit of the principal chiral model on the supersphere. In
these cases, one way to test a duality is to compute certain quantities,
e.g. some boundary spectra, at this solvable point and perform the
perturbation to other points in the moduli space exactly
\cite{Quella:2007sg}. This method has successfully been applied to the
supersphere/Gross-Neveu correspondence \cite{Mitev:2008yt}. The question
remains how to actually prove such a correspondence. The case $N=0$ in the
Gross-Neveu-supersphere duality is the well-known correspondence between
the O(2) Gross-Neveu model, that is the massless Thirring model, and a
free boson on the circle, i.e. bosonization \cite{Coleman:1974bu}.
Unfortunately, the proof does not generalize straightforwardly, but still
we believe that bosonization techniques will turn out to be crucial in
understanding the correspondence.

If there is a simpler model at hand, it is a good idea to study it in
detail to gain insight and to establish techniques for the more
complicated models. This leads us to the \GL-symplectic fermion
correspondence. The \GL\ WZNW model is probably the best understood CFT
with supersymmetric target space that is not free. There exists another
CFT with \GL\ current symmetry \cite{LeClair:2007aj}, which was used to
study the spin quantum Hall transition. This CFT is constructed from the
OSp(2$|$2) Gross-Neveu model at the free point via bosonization and it
automatically has \GL\ symmetry since the OSp(2$|$2) Gross-Neveu model is
constructed from a spin one half vector transforming in the adjoint
representation of \GL. This model consists of two free scalars and a set
of symplectic fermions. The symplectic fermions were first analyzed in
detail in~\cite{Kausch:1995py,Kausch:2000fu}. The first part of this note
is devoted to showing the correspondence between these models. The
technique we use is based on bosonization, but in addition we use the
affine currents as a guideline which we hope is also useful for other
Gross-Neveu-like models.

The correspondence, we find, is remarkable in its own right since the \GL\
WZNW model is an interacting theory, while the corresponding model is
free, and the bosons are completely decoupled from the fermions. The
non-triviality is hidden in the vertex operators, i.e. the \GL\ vertex
operators in the free description contain twist fields of the fermions and
it turns out that the computation of bulk correlation functions in both
descriptions is of a similar complexity. Still our method provides a new
approach to WZNW models on Lie supergroups. So far, the models have been
investigated either algebraically \cite{Rozansky:1992rx} or in terms of
fermionic ghost systems
\cite{Schomerus:2005bf,Gotz:2006qp,Quella:2007hr,Creutzig:2008ek}.
Hopefully, there exist generalizations of our approach to other Lie
supergroups leading to a better understanding of them.

In the second part of this note, we apply the correspondence to branes in
\GL. For the understanding of Cardy boundary states, the free description
is better adapted than the original one. \GL\ possesses two classes of
branes. One of them, the so-called untwisted branes which geometrically
describe superconjugacy classes in the supergroup manifold, have been
studied in detail in \cite{Creutzig:2007jy}. It was found that amplitudes
of boundary states satisfy Cardy conditions~\cite{Cardy:1989ir} and that
they agree with fusion, as expected from experience with rational
CFT~\cite{Gaberdiel:2002qa} and also logarithmic
CFT~\cite{Gaberdiel:2007jv}. The second class of branes contains just one
volume-filling brane. This brane has been investigated in
\cite{Creutzig:2008ek}, i.e. its correlation functions have been computed,
but also the boundary state has been constructed and tested. For a
complete description of boundary states one still needs to understand
amplitudes of strings stretching between a twisted and an untwisted brane.
In our new description this can be done straightforwardly. The final
result is that Cardy conditions are still satisfied, and essentially all
known results of branes on Lie groups carry over to the Lie supergroup
\GL.

The structure of this note is as follows. In section 2 we verify the
correspondence in detail. We explain how the currents are used as a
guideline to prove the correspondence, and we check the correspondence by
comparing correlation functions. Section 3 gives a detailed discussion of
boundary states in the symplectic fermions, including twist fields. In
section 4 we apply the results of the two previous sections to complete
the discussion of \GL\ boundary states.

\section{The \GL-symplectic fermion correspondence}

In this section we will set up the notation and show the relation between
the \GL\ WZNW model and the free scalars and symplectic fermions. Finally,
we will comment on the bulk correlation functions.

\subsection{The \GL\ WZNW model}

Our starting point for the relation between the \GL\ WZNW model and the
free theory will be the first order action for \GL\ found
in~\cite{Schomerus:2005bf}. To set up the notation used in this paper we
recall a few facts about the \gl\ superalgebra. It is generated by two
bosonic elements $E, N$ and two fermionic $\psi^{\pm}$ which have the
following non-zero (anti)commutator relations
\begin{align}\label{eq:commu}
    [N,\psi^{\pm}]=\pm\psi^{\pm},\quad \{\psi^-,\psi^+\}=E.
\end{align}
Further, we have a family of supersymmetric bilinear forms, but below we
will always work with
\begin{align}\label{eq:supertrace}
    \str(NE)=\str(\psi^+\psi^-)=-1.
\end{align}
For the \GL\ supergroup we choose a Gauss-like decomposition of the form
$$ g \ = \ e^{c_- \psi^-} \, e^{XE + YN} \, e^{-c_+
\psi^+}. $$
The WZNW model thus has two bosonic fields $X(z,\bar z),Y(z, \bar z)$ and two fermionic
fields $c_\pm(z, \bar z)$, and the action takes the form
\begin{equation}\label{eq:SWZW}
    \begin{split}
S_{\text{WZNW}}[g(X,Y,c_\pm)]\ &= \frac{k}{4 \pi} \int_{\Sigma} d ^2z
 \langle g^{-1} \partial g , g^{-1} \bar \partial g \rangle
 + \frac{k}{2 4 \pi} \int_{B} \langle g^{-1} d g ,
  [g^{-1} d g , g^{-1} d g ] \rangle\\ &=\  \frac{k}{4\pi }\int_\Sigma
d^2z\
  \left( -\del X\bar{\del}Y-\del
  Y\bar{\del}X+2e^{Y}\del\cp\bar{\del}\cm\right)\, ,
  \end{split}
\end{equation}
where $k$ is the level. Variation of the action leads to the usual bulk
equations of motion \cite{Creutzig:2007jy}.

The holomorphic current of the \GL\ WZNW model is in our notation given by
$k\del g g^{-1}$. The components corresponding to the generators are
\begin{align}
    J^E &= -k\del Y, &
    J^N &= -k\del X +k\cm \del\cp \, e^{Y} \ ,\nonumber \\[2mm]
    J^- &= ke^{Y}\del\cp, &
    J^+  &= -k\del\cm -k\cm \del Y \ ,\label{eq:holomorphiccurrents}
\end{align}
Similarly, for the anti-holomorphic current $-kg^{-1}\delbar g$ the
components are
\begin{align}
        \bar{J}^E &= k\bar{\del} Y,  &
        \bar{J}^N &= k\bar{\del} X -k \bar{\del}\cm\, \cp\, e^{Y} \ , \nonumber
        \\[2mm]
            \bar{J}^+&= ke^{Y}\bar{\del}\cm, &
        \bar{J}^-&= -k\bar{\del}\cp-k\cp\bar{\del} Y \ .\label{eq:antiholomorphiccurrents}
\end{align}

Let us also mention that the modes of this affine algebra satisfy
\begin{align}\label{eq:affine}
[J^E_n,J^N_m]=-km\de_{n+m},\quad[J^N_n, J^{\pm}_m]=\pm J^{\pm}_{n+m},\quad \{J^-_n,J^+_m\}=J^E_{n+m}+km\de_{n+m},
\end{align}
where we note that the modes can be rescaled such that the algebra is
independent of the level $k$. Equation~\eqref{eq:affine} corresponds to the
OPE
\begin{equation}\label{eq:currentope}
    J^A(z) J^B (w)\sim -k\frac{\str(AB)}{(z-w)^2}+\frac{[A,B\}}{z-w}.
\end{equation}

\subsection{First order formulation}

Following~\cite{Schomerus:2005bf} we will now pass to a first order
formalism by introducing two additional fermionic auxiliary fields $b_\pm$
of weight $\Delta(b_\pm) = 1$. Naively, the action would be
\begin{equation}
        \frac{1}{4\pi}\int_\Sigma d^2z\ \left(- k\del X\bar{\del}Y-
        k\del Y\bar{\del}X+2\bp\del\cp+2\bm\bar{\del}\cm+\frac{2}{k}e^{-Y}\bm\bp\right).
\end{equation}
This reduces to~\eqref{eq:SWZW} if we integrate out $b_\pm$ using their
equations of motion
\begin{align}\label{eq:polbb}
   b_- = k \del c_+ \exp
Y,\qquad b_+ = - k \bar \del c_- \exp Y.
\end{align}
However, we get a quantum correction in going from the \GL\ invariant
measure used for the action in~\eqref{eq:SWZW} to the free-field measure
$\mathcal{D}X\mathcal{D}Y
      \mathcal{D}\cm\mathcal{D}\cp\mathcal{D}\bm\mathcal{D}\bp$ that we
want to use for our first order formalism. In analogy with
\cite{Gerasimov:1990fi} the correction is
\begin{align}\label{eq:polmeas}
    \ln\det\left(|\rho|^{-2}e^{-Y}\del e^{Y}\bar \del\right)=\frac{1}{4\pi}\int d^2 z\left(\del Y\bar\del Y+\frac{1}{4}\sqrt{G}\mathcal{R}Y\right).
\end{align}
Here $G$ is the determinant of the world-sheet metric and $\mathcal{R}$
its Gaussian curvature. $|\rho|^2$ is the metric and we have the relation
$\sqrt{G}\mathcal{R} \ = \ 4 \del\bar{\del}\log |\rho|^2$. We thus get a
correction to the kinetic term and a background charge for $Y$. The first
order action including the correction is
\begin{multline}\label{eq:pol1st}
    S(X,Y,b_{\pm},c_\pm) = \frac{1}{4\pi}\int_\Sigma
d^2z \bigg(-k\del X\bar{\del}Y-k\del
  Y\bar{\del}X+\del
  Y\bar{\del}Y+\frac{1}{4}\sqrt{G}\mathcal{R}Y \\ +2\cp\del\bp+2\cm\bar{\del}\bm+\frac{2}{k}e^{-Y}\bm\bp\bigg).
\end{multline}

We also get a quantum correction to the current. This will happen where we
have to choose a normal ordering of the terms in the
current~\eqref{eq:holomorphiccurrents}. We fix this by demanding that the
currents obey the OPEs~\eqref{eq:currentope}. Indeed, we have to add $\del
Y$ to $J^N$ to ensure that it has a regular OPE with itself. Thus the
holomorphic currents in the free field formalism are
\begin{align*}
    J^E &= -k\del Y ,  &
    J^N &= -k\del X +\cm b_-+\del Y \ , \\[2mm]
    J^- &= b_-,  &
    J^+ &=  -k\del\cm -k\cm \del Y \ ,
\end{align*}
where we here and in the following suppress the normal ordering. We get
similar expressions for the anti-holomorphic currents.

\subsection{The correspondence}

If we integrate out $b_\pm$ in~\eqref{eq:pol1st} we would simply obtain
the \GL\ WZNW model. We will now show that if we instead bosonize the $bc$
system to obtain a system of three scalars, it is possible to perform a
field redefinition such that one of the scalars decouples. We can then
return to a new $b'c'$ formalism and integrate out $b'_\pm$ to arrive at a
decoupled theory of two scalars and a set of symplectic fermions.

In this process the current becomes more symmetric and simple. It can be
seen as a guideline for which transformations to perform and we will
therefore explicitly follow the transformation of the current in each
step.

We will start by only discussing the transformation of the action and the
current. The map of the vertex operators will be determined in the next
subsection.

To begin we bosonize the $bc$ system in~\eqref{eq:pol1st} in the standard
way~\cite{Friedan:1985ge}
\begin{align}
    c_\pm =e^{\rho^{R,L}},&\qquad b_\pm=e^{-\rho^{R,L}},\nonumber \\
    \cp\del\bp+\cm\bar{\del}\bm &= -\frac{1}{2}\del\rho\bar\del\rho+\frac{1}{8}\sqrt{G}\mathcal{R}\rho,\nonumber \\
    b_- c_-&=-\del\rho^L, \label{eq:polbos}
\end{align}
where we denote left and right components of scalars by superscripts
$L,R$. In the currents we likewise have to introduce left and right
indices and the holomorphic currents then become
\begin{align}
    J^E &=\ -k\del Y^L,  &
    J^N &=\ -k\del X^L +\del\rho^L+\del Y^L \ ,\nonumber \\[2mm]
    J^- &=\ e^{-\rho^L},  &
    J^+ &= \ -k\del(\rho^L + Y^L)e^{\rho^L} \ ,
\end{align}
and the action is
\begin{equation}\label{eq:polbosact}
\begin{split}
    S(X,Y,b_{\pm},c_\pm) \ = \ \frac{1}{4\pi}\int_\Sigma
                  d^2z \Bigl(-k\del X\bar{\del}Y&-k\del  Y\bar{\del}X+\del   Y\bar{\del}Y+\Bigr. \\
                         \Bigl.&-\del\rho\bar\del\rho+\frac{1}{4}\sqrt{G}\mathcal{R}(Y+\rho)+\frac{2}{k}e^{-Y-\rho}\Bigr)\, .\\
  \end{split}
\end{equation}

We observe, both from the current and the action, that it is very
natural to go to variables $Y, Z, \rho'$ where
\begin{align}\label{eq:polphi1phi2}
    \rho'=Y+\rho,\qquad Z=kX-\rho-Y=kX-\rho'.
\end{align}
The currents and the action in these variables are
\begin{align}
    J^E &= -k\del Y^L,& J^N&=-\del Z^L,\nonumber \\
    J^- &= e^{Y^L-\rho'^L},& J^+ &= -k\del\rho'^Le^{\rho'^L-Y^L},
\end{align}
\begin{align}
    S(X,Y,b_{\pm},c_\pm) = \frac{1}{4\pi}\int_\Sigma
d^2z \left(-\del Z\bar{\del}Y-\del
  Y\bar{\del}Z-\del\rho'\bar\del\rho'+\frac{1}{4}\sqrt{G}\mathcal{R}\rho'+\frac{2}{k}e^{-\rho'}\right)\, .\label{eq:poltransact}
\end{align}
Hence we got a theory of two scalars decoupled from a Coulomb gas with
screening charge. For calculation of correlation functions this is a very
efficient formulation of the theory. We will, however, go one step further
and rewrite the screened Coulomb gas in terms of symplectic fermions.

We thus return to a $b'c'$ system using again~\eqref{eq:polbos}, but now
for the field $\rho'$. This gives us the following simple expressions
\begin{align}
    J^E &= -k\del Y^L,& J^N &=-\del Z^L,\nonumber \\
    J^- &= e^{Y^L}b'_-, & J^+ &= -ke^{-Y^L}\del c'_-\, ,
\end{align}
\begin{align}
    S(X,Y,b'_{\pm},c'_\pm) = \frac{1}{4\pi}\int_\Sigma
d^2z \left(-\del Z\bar{\del}Y-\del
  Y\bar{\del}Z+2\cp'\del\bp'+2\cm'\bar{\del}\bm'+\frac{2}{k}b'_- b'_+\right)\, .\label{eq:polfinal}
\end{align}

We can now integrate out the fields $b'_\pm$ getting the equations of
motion
\begin{align}\label{eq:polbs}
    b'_+=-k\bar\del c'_-,\quad b'_-=k\del c'_+\, ,
\end{align}
and arrive at
\begin{align}
    S(X,Y,c_\pm) = \frac{1}{4\pi}\int_\Sigma
d^2z \left(-\del Z\bar{\del}Y-\del
  Y\bar{\del}Z+2k\del\cp'\bar{\del}\cm'\right)\, .\label{eq:polfinalsymplecticfermion1}
\end{align}
Of course, we have to be careful when the vertex operators depend on $b'$. As
we will see below, the vertex operators for typical representations will
be twist operators which we interpret as not containing $b$.

To remove the dependence on the level $k$ in the action we introduce
$\chi^a$ by
\begin{align}\label{eq:rescaling}
    \sqrt{k}c'_+=\chi^1,\quad \sqrt{k}c'_-=\chi^2,
\end{align}
and the currents and action are then
\begin{align}
    J^E &= -k\del Y^L,& J^N &=-\del Z^L,\nonumber \\
    J^- &= \sqrt{k}e^{Y^L}\del \chi^1, & J^+ &= -\sqrt{k}e^{-Y^L}\del \chi^2,\label{eq:finalcurrent}
\end{align}
\begin{align}
    S(X,Y,\chi^a) = \frac{1}{4\pi}\int_\Sigma
d^2z \left(-\del Z\bar{\del}Y-\del
  Y\bar{\del}Z+\ep_{ab}\del\chi^a\bar{\del}\chi^b\right)\, .\label{eq:polfinalsymplecticfermion2}
\end{align}
where the anti-symmetric symbol is defined by $\ep_{12}=-\ep_{21}=1$. This
gives the OPEs
\begin{align}
    \chi^a(z,\bar z)\chi^b(w,\bar w)&\sim -\ep^{ab}\ln\abs{z-w}^2,\nonumber \\
    Z(z,\bar z) Y(w,\bar w)&\sim\ln\abs{z-w}^2.\label{eq:polopesympl}
\end{align}
where $\ep^{12}=-1$. This is the action and current that was constructed
in~\cite{LeClair:2007aj}. In that reference it was also found that the
action has an enlarged OSp$(2|2)$ symmetry.

For future reference, let us sum up the correspondence between the
symplectic fermions and the underlying $b',c'$ system. We have
\begin{align}
    \delbar\chi^1&=\sqrt{k}\delbar c'_+, & \delbar\chi^2&=\sqrt{k}\delbar c'_-=-\frac{1}{\sqrt{k}}b'_+,\nn \\
    \del\chi^1&=\sqrt{k}\del c'_+=\frac{1}{\sqrt{k}}b'_-, & \del\chi^2&=\sqrt{k}\del c'_-,\label{eq:polderivdict}
\end{align}
which will be useful in the next section where we study what happens to
the vertex operators.

\subsection{Mapping of the vertex operators}

We now consider the mapping of the \GL\ vertex operators under the
transformation that we found in the last subsection. The basis of vertex
operators to be used with the first order action~\eqref{eq:pol1st} were
found in~\cite{Schomerus:2005bf} by a minisuperspace analysis.  We will
here use the notation of~\cite{Creutzig:2008ek} and write the operators as
\begin{equation}\label{eq:polvertexope1}
    V_{\<-e,-n+1\>}=\normord{e^{eX+nY}}\begin{pmatrix}
                           1 & c_-\\
                           c_+ & c_-c_+ \\
                         \end{pmatrix},
\end{equation}
and the conformal dimension is
\begin{equation}\label{eq:polconfdim}
    \De_{(e,n)}=\frac{e}{2k}(2n-1+\frac{e}{k}).
\end{equation}
For $e\neq mk$, where $m$ is an integer, the columns of this matrix will
correspond to the two-dimensional representation $\<-e,-n+1\>$ for the left-moving currents while the rows correspond to
the representation $\<e,n\>$ under the right-moving currents.

Let us first consider the transformation giving us~\eqref{eq:poltransact}:
\begin{gather}
    X=\frac{1}{k}(\rho'+Z),\nonumber\\
    c_-=e^{\rho'^L_1-Y^L},\quad b_-=e^{-\rho'^L_1+Y^L}.\label{eq:poltranstocol}
\end{gather}
This maps the vertex operators to
\begin{equation}\label{eq:polvertexope2}
    V_{\<-e,-n+1\>}=\normord{e^{\frac{e}{k}\rho'+\frac{e}{k}Z+nY}\begin{pmatrix}
                           1 & e^{\rho'^L-Y^L}\\
                           e^{\rho'^R-Y^R} & e^{\rho'-Y} \\
                         \end{pmatrix}}.
\end{equation}
Here we generally split scalar fields into the left and right handed part
as $\rho'=\rho'^L+\rho'^R$. Some comments are in order here: Firstly,
rather than thinking of e.g. $c_-$ in~\eqref{eq:polvertexope1} as a
function to be evaluated under the path integral, we have here used
bosonization and will think about the vertex operators in the operator
formalism. This means that $c_-$ is a holomorphic operator. Secondly, for
the $YZ$ system the vertex operators are
\begin{equation}\label{eq:polvertexope2.1}
V_{\<-e,-n+1\>}^{\textrm{B}}=\begin{pmatrix}
  \normord{e^{\frac{e}{k}Z+nY}} & \normord{e^{\frac{e}{k}Z+(n-1)Y^L+nY^R}} \\
  \normord{e^{\frac{e}{k}Z+nY^L+(n-1)Y^R}} & \normord{e^{\frac{e}{k}Z+(n-1)Y}} \\
\end{pmatrix},
\end{equation}
whereas for the $\rho'$ system they are
\begin{equation}\label{eq:polvertexope2.2}
    V_{\<-e,-n+1\>}^{\textrm{F}}=\begin{pmatrix}
    \normord{e^{\frac{e}{k}\rho'}} & \normord{e^{(\frac{e}{k}+1)\rho'^L+\frac{e}{k}\rho'^R}}\\
    \normord{e^{\frac{e}{k}\rho'^L+(\frac{e}{k}+1)\rho'^R}} & \normord{e^{(\frac{e}{k}+1)\rho'}} \\
                         \end{pmatrix}.
\end{equation}
Thus in the off-diagonal terms, the splitting into holomorphic and
anti-holomorphic parts means that the correlation functions calculated in
respectively the $YZ$ system and the $\rho'$ system are not separately
real, but only the combined correlation function can be expressed in the
absolute values of the insertions $z_i$. Also, we see that around the
off-diagonal terms in the operator~\eqref{eq:polvertexope2.1} the field
$Z$ gets an additive twist. The overall twist vanishes due to charge
conservation for $Y$.

Since $\rho'$ now appears with non-integer momenta, we see that in going to
the $b',c'$ system with action~\eqref{eq:polfinal} we get twist operators.
Precisely, the vertex operator~\eqref{eq:polvertexope2.2} maps into
\begin{align}\label{eq:polvertexopebc}
V_{\<-e,-n+1\>}^{\textrm{F}}=\begin{pmatrix}
  \tilde{\mu}^L_{e/k}\tilde{\mu}^R_{e/k} & \tilde{\mu}^L_{e/k+1}\tilde{\mu}^R_{e/k} \\
  \tilde{\mu}^L_{e/k}\tilde{\mu}^R_{e/k+1} & \tilde{\mu}^L_{e/k+1}\tilde{\mu}^R_{e/k+1} \\
\end{pmatrix},
\end{align}
where the twist states are defined by
\begin{align}\label{eq:poltwist}
    c'_-(e^{2\pi i}z)\tilde{\mu}^L_{\la}(0)=e^{2\pi i\la}\tilde{\mu}^L_{\la}(0).
\end{align}
This is solved by
\begin{align}\label{eq:poltwisttoexp}
    \tilde{\mu}^L_{\la}\equiv\normord{e^{\la\rho'^L}},
\end{align}
but only uniquely in $\la$ modulo integers and, naturally, up to a
normalisation. The conformal dimension is $-\half\la(1-\la)$ so the ground
states have $0<\la<1$. We can step $\la$ up and down with respectively
$c'_-$ and $b'_-$ e.g.
\begin{align}\label{eq:poltwist2}
    c'_-(z)\tilde{\mu}^L_{\la}(0)\sim\frac{1}{z^{-\la}}\tilde{\mu}^L_{\la+1}(0).
\end{align}
Also note that
\begin{align}
    \tilde{\mu}^R_{\la}\equiv\normord{e^{\la\rho'^R}},
\end{align}
fulfills
\begin{align}\label{eq:poltwist3}
    c'_+(e^{-2\pi i}{\bar z})\tilde{\mu}^R_{\la}(0)=e^{-2\pi i\la}\tilde{\mu}^R_{\la}(0).
\end{align}
Since $\tilde{\mu}^R_\la$ gives opposite transformations compared to the
holomorphic operator $\tilde{\mu}^L_\la$, but has the same dimension
$-\half\la(1-\la)$, it in many ways compares to $\tilde{\mu}^{L}_{1-\la}$.

To obtain the symplectic fermions requires integrating out $b'$. This
means that the anti-holomorphic part of $c'_-$ is non-trivial in the OPEs.
As an example, $c'_+$ and $c'_-$ with
action~\eqref{eq:polfinalsymplecticfermion1} have a singular OPE that is
$\sim\frac{1}{k}\ln\abs{z-w}^2$. However, using
equations~\eqref{eq:polderivdict} we get the mapping of $\del c_-'$ and
$b_-'$ to the holomorphic operators $\del\chi^2$ and $\del\chi^1$.
Likewise, $\delbar c_+'$ and $b_+'$ will correspond to the
anti-holomorphic operators $\delbar\chi^1$ and $\delbar\chi^2$.

One has to be careful since we in principle can not integrate out $b'$
when the vertex operators depend on $b'_-b'_+$. However, for the twist
operators it seems plausible since, at least for $\la>0$, we can naively
think of $\mu_\la$ as $c'^\la$. To check this we will in the next section
compare the correlation functions to the already known calculation for the
symplectic fermions. The twist fields in the $b',c'$ system then directly
translates into twist fields of the symplectic fermions. The symplectic
fermion twist fields are defined by~\cite{Kausch:2000fu}
\begin{align}
    \chi^1(e^{2\pi i}z)\mu_\la(0)&=e^{-2\pi i\la}\chi^1(z)\mu_\la(0), & \chi^2(e^{2\pi i}z)\mu_\la(0)&=e^{2\pi i\la}\chi^2(z)\mu_\la(0),\nonumber\\
    \bar \chi^1(e^{-2\pi i} \bar z)\mu_\la(0)&=e^{-2\pi i\la}\bar \chi^1(\bar z)\mu_\la(0), & \bar \chi^2(e^{-2\pi i}\bar z)\mu_\la(0)&=e^{2\pi i\la}\bar \chi^2(\bar z)\mu_\la(0),\label{eq:polsympltwist}
\end{align}
where $\chi^1$ and $\chi^2$ has to transform oppositely to give a symmetry
of the Lagrangian. Here we have split the symplectic fermions into their
chiral and anti-chiral parts
$\chi^a(z,\bar{z})=\chi^a(z)+\bar{\chi}^a(\bar{z})$. The anti-holomorphic
part must transform in the same way under $\bar{z}\mapsto e^{-2\pi
i}\bar{z}$, but importantly $\la$ can differ by an integer between the
holomorphic and anti-holomorphic sector. The
condition~\eqref{eq:polsympltwist} is fulfilled by
$\tilde{\mu}_\la^L\tilde{\mu}_\la^R$ and the other operators
in~\eqref{eq:polvertexopebc}. However, we have done the
rescaling~\eqref{eq:rescaling} so if we think of the twist operator as
$(c'_-)^\la$ we should choose the following normalisation:
\begin{align}\label{eq:newtwistfield}
    \mu^L_\la=\sqrt{k}^\la\tilde{\mu}^L_\la=\sqrt{k}^\la\normord{e^{\la\rho'^L}},
\end{align}
and similarly for the anti-holomorphic part. Thus the vertex
operator~\eqref{eq:polvertexopebc} maps into
\begin{align}\label{eq:polvertexopesymplectic}
V_{\<-e,-n+1\>}^{\textrm{F}}\mapsto k^{-\frac{e}{k}}\begin{pmatrix}
  {\mu}^L_{e/k}{\mu}^R_{e/k} & \frac{1}{\sqrt{k}}{\mu}^L_{e/k+1}{\mu}^R_{e/k} \\
  \frac{1}{\sqrt{k}}{\mu}^L_{e/k}{\mu}^R_{e/k+1} & \frac{1}{k}{\mu}^L_{e/k+1}{\mu}^R_{e/k+1} \\
\end{pmatrix}.
\end{align}
A notation with splitting into left and right part, like in the $b'c'$
system, turns out to be useful. The twist values can be stepped up and
down using the following OPEs:
\begin{align}\label{eq:polsympltwistope1}
    \del\chi^1(z)\mu^L_\la(0)&\sim\frac{1}{z^\la}\mu^L_{\la-1}(0),\quad \del\chi^2(z)\mu^L_{\la}(0)\sim\frac{\la}{z^{1-\la}}\mu^L_{\la+1}(0),
\end{align}
and correspondingly
\begin{align}\label{eq:polsympltwistope2}
    \delbar\bar \chi^1(\bar z)\mu^R_\la(0)&\sim\frac{\la}{\bar z^{1-\la}}\mu^R_{\la+1}(0), \quad \delbar\bar \chi^2(\bar z)\bar \mu^R_\la(0)\sim-\frac{1}{\bar z^{\la}}\mu^R_{\la-1}(0).
\end{align}
We note here again that up to a sign the anti-holomorphic side is
understood by seeing $\mu^R_\la$ as $\mu^L_{1-\la}$.

To conclude the total vertex operator $V_{\<-e,-n+1\>}$ in the $YZ$ and
symplectic fermion system with
action~\eqref{eq:polfinalsymplecticfermion2} takes the form
\begin{align}\label{eq:finalvertexoperator}
   V_{\<-e,-n+1\>}\mapsto k^{-\frac{e}{k}} \begin{pmatrix}
  \normord{e^{\frac{e}{k}Z+nY}}{\mu}^L_{e/k}{\mu}^R_{e/k} & \frac{1}{\sqrt{k}}\normord{e^{\frac{e}{k}Z+(n-1)Y^L+nY^R}}{\mu}^L_{e/k+1}{\mu}^R_{e/k} \\
  \frac{1}{\sqrt{k}}\normord{e^{\frac{e}{k}Z+nY^L+(n-1)Y^R}}{\mu}^L_{e/k}{\mu}^R_{e/k+1} & \frac{1}{k}\normord{e^{\frac{e}{k}Z+(n-1)Y}}{\mu}^L_{e/k+1}{\mu}^R_{e/k+1} \\
\end{pmatrix}
\end{align}
We note that equations~\eqref{eq:polsympltwistope1} can be used to check
that the columns of this operator transform in the \mbox{$\<-e,-n+1\>$}
representation of \GL\ under the left-moving
currents~\eqref{eq:finalcurrent}. These operators are indeed close to the
operators found in~\cite{LeClair:2007aj}. Let us now check the operators
in correlation functions.

\subsection{Bulk correlation functions}

We will now compare the correlation functions of the primary
fields~\eqref{eq:polvertexope1} obtained in the \GL\ model to the
calculations done for the symplectic fermions in~\cite{Kausch:2000fu}. The
similarity was already noted in~\cite{Schomerus:2005bf}.

Let us first note that from equations~\eqref{eq:polvertexope2.1}
and~\eqref{eq:polvertexope2.2} the vertex
operators~\eqref{eq:polvertexope1} in the $Y,Z,\rho'$
picture~\eqref{eq:poltransact} takes the form
\begin{align}\label{eq:vertexwithindices}
    {V_{\<-e,-n+1\>}}_\si^{\bar\si}=\normord{e^{\frac{e}{k}Z+(n-\si)Y^L+(n-\bar\si)Y^R}e^{(\frac{e}{k}+\si)\rho'^L+(\frac{e}{k}+\bar\si)\rho'^R}},
\end{align}
where $\si,\bar\si\in\{0,1\}$ labels respectively the columns and the
rows.

We consider the three-point function
\begin{align}\label{eq:3pt1}
    A=\expect{{V_{\<-e_1,-n_1+1\>}}_{\si_1}^{\bar\si_1}(z_1){V_{\<-e_2,-n_2+1\>}}_{\si_2}^{\bar\si_2}(z_2){V_{\<-e_3,-n_3+1\>}}_{\si_3}^{\bar\si_3}(z_3)}.
\end{align}
The correlation function splits into a $YZ$ and a $\rho'$ part,
$A=A^{\textrm{B}}A^{\textrm{F}}$. The $YZ$ part is easily evaluated to be
\begin{multline}\label{eq:3pt2}
    A^{\textrm{B}}=\delta\big(\sum_i \frac{e_i}{k}\big)\delta\big(\sum_i(n_i-\si_i)\big)\delta\big(\sum_i(n_i-\bar\si_i)\big)\\
    \times\prod_{i<j}(z_i-z_j)^{\frac{e_i}{k}(n_j-\si_j)+\frac{e_j}{k}(n_i-\si_i)}(\bar z_i-\bar
    z_j)^{\frac{e_i}{k}(n_j-\bar\si_j)+\frac{e_j}{k}(n_i-\bar\si_i)},
\end{multline}
where the indices run from 1 to 3. The $\de$-functions follow directly
from the $J^E$ and $J^N$ currents. The $\rho'$ part is also easily
evaluated. Here one has to remember that the overall $\rho'$ charge has to
sum to one due to the background charge of $\rho'$. This means that we can
maximally have two insertions of the interaction term of the
action~\eqref{eq:poltransact}. However, as was commented
in~\cite{Schomerus:2005bf}, the part with two interaction terms vanish.
The part with one interaction term is calculated using the Dotsenko-Fateev
like integral used in~\cite{Schomerus:2005bf}. We get
\begin{align}
    A^{\textrm{F}}=\delta\big(\sum_i\si_i -1\big)\delta\big(\sum_i\bar\si_i -1\big)&\prod_{i<j}(z_i-z_j)^{(\frac{e_i}{k}+\si_i)(\frac{e_j}{k}+\si_j)}(\bar z_i-\bar z_j)^{(\frac{e_i}{k}+\bar \si_i)(\frac{e_j}{k}+\bar \si_j)}\nonumber \\
    -\frac{1}{k}\delta\big(\sum_i\si_i -2\big)\delta\big(\sum_i\bar\si_i -2\big)&(-1)^{\si_3+\bar\si_3}\frac{\Ga(1-\frac{e_1}{k}-\si_1)\Ga(1-\frac{e_2}{k}-\si_2)\Ga(1-\frac{e_3}{k}-\bar\si_3)}{\Ga(\frac{e_3}{k}+\si_3)\Ga(\frac{e_1}{k}+\bar \si_1)\Ga(\frac{e_2}{k}+\bar\si_2)} \nonumber\\
    \times &\prod_{i<j}(z_i-z_j)^{(\frac{e_i}{k}+\si_i-1)(\frac{e_j}{k}+\si_j-1)}(\bar z_i-\bar z_j)^{(\frac{e_i}{k}+\bar \si_i-1)(\frac{e_j}{k}+\bar \si_j-1)}\label{eq:3pt3},
\end{align}
where the first part is for no interaction term and the second part for
one interaction term. We have here used that $\sum_i e_i=0$ due to the
delta-function from the $YZ$ part of the correlation function
in~\eqref{eq:3pt2}.

If we combine the two parts in~\eqref{eq:3pt2} and~\eqref{eq:3pt3} the
symmetry between the holomorphic and anti-holomorphic sector is restored
and we arrive at
\begin{multline}\label{eq:3pt4}
    A=\delta\big(\sum_i \frac{e_i}{k}\big)\delta\big(\sum_i(n_i-\si_i)\big)\delta\big(\sum_i(n_i-\bar\si_i)\big)\\
    \bigg(\delta\big(\sum_i\si_i -1\big)\delta\big(\sum_i\bar\si_i -1\big)\prod_{i<j}\abs{z_i-z_j}^{2\frac{e_i}{k}n_j+2\frac{e_j}{k}n_i+2\frac{e_ie_j}{k^2}} \\
    -\frac{1}{k}\delta\big(\sum_i\si_i -2\big)\delta\big(\sum_i\bar\si_i -2\big)(-1)^{\si_3+\bar\si_3}\frac{\Ga(1-\frac{e_1}{k}-\si_1)\Ga(1-\frac{e_2}{k}-\si_2)\Ga(1-\frac{e_3}{k}-\bar\si_3)}{\Ga(\frac{e_3}{k}+\si_3)\Ga(\frac{e_1}{k}+\bar \si_1)\Ga(\frac{e_2}{k}+\bar\si_2)} \\
    \times \prod_{i<j}\abs{z_i-z_j}^{2\frac{e_i}{k}(n_j-1)+2\frac{e_j}{k}(n_i-1)+2\frac{e_ie_j}{k^2}}
    \bigg),
\end{multline}
as was derived in~\cite{Schomerus:2005bf}. This indeed supports the
validity of our decoupling of the \GL\ WZNW model into a set of free
scalars and the $\rho'$ system with action~\eqref{eq:poltransact}. The
result may not look local, e.g. does not seem to be symmetric in
interchanging operator 2 and 3, due to the asymmetric-looking $\Ga$
functions. However, these can be rewritten in the following symmetric form
\begin{align}\label{eq:3pt4.5}
    (-1)^{\si_3+\bar\si_3}\frac{\Ga(1-\frac{e_1}{k}-\si_1)\Ga(1-\frac{e_2}{k}-\si_2)\Ga(1-\frac{e_3}{k}-\bar\si_3)}{\Ga(\frac{e_3}{k}+\si_3)\Ga(\frac{e_1}{k}+\bar \si_1)\Ga(\frac{e_2}{k}+\bar\si_2)}=\prod_i\frac{\Ga(1-\frac{e_i}{k})}{\Ga(\frac{e_i}{k})}\left(\frac{-e_i}{k}\right)^{-\si_i-\bar\si_i}.
\end{align}

As we see from the result~\eqref{eq:3pt4} one has to be careful in the
limit when $e_i$ is an integer multiple of $k$. As was shown
in~\cite{Schomerus:2005bf} this gives logarithmic correlation functions.
For now let us not consider these limits. Thus we get genuine twist
operators when going to the symplectic fermions and the twists are
$\la_i=e_i/k+\si_i$ in the holomorphic sector and
$\bar\la_i=e_i+\bar\si_i$ in the anti-holomorphic sector when we compare
equation~\eqref{eq:vertexwithindices}
with~\eqref{eq:polvertexopesymplectic}. As we see from the vertex
operators in~\eqref{eq:polvertexopesymplectic}, the results that we expect
from the symplectic fermions to comply with correlation
function~\eqref{eq:3pt3} are
\begin{multline}\label{eq:3pt5}
    \expect{\mu^L_{\la_1}(z_1)\mu^R_{\bar \la_1}(\bar z_1)\mu^L_{\la_2}(z_2)\mu^R_{\bar \la_2}(\bar z_2)\mu^L_{\la_3}(z_3)\mu^R_{\bar \la_3}(\bar z_3)}_{\mathrm{SF}}=\prod_{i<j}(z_i-z_j)^{\la_i\la_j}(\bar z_i-\bar z_j)^{\bar \la_i\bar
    \la_j}\\
    \quad\textrm{for }\sum_i\la_i=\sum_i\bar\la_i=1,
\end{multline}
and
\begin{multline}\label{eq:3pt6}
    \expect{\mu^L_{\la_1}(z_1)\mu^R_{\bar \la_1}(\bar z_1)\mu^L_{\la_2}(z_2)\mu^R_{\bar \la_2}(\bar z_2)\mu^L_{\la_3}(z_3)\mu^R_{\bar \la_3}(\bar z_3)}_{\mathrm{SF}}\\
    =-(-1)^{\la_3-\bar\la_3}\frac{\Ga(\la_1^*)\Ga(\la_2^*)\Ga(\bar\la_3^*)}{\Ga(\bar\la_1)\Ga(\bar\la_2)\Ga(\la_3)}\prod_{i<j}(z_i-z_j)^{\la^*_i\la^*_j}(\bar z_i-\bar z_j)^{\bar \la^*_i\bar
    \la^*_j}\quad\textrm{for }\sum_i\la_i=\sum_i\bar\la_i=2,
\end{multline}
where $\la^*=1-\la$ and the subscript SF means that the expectation value
is calculated using the symplectic fermion part of the
action~\eqref{eq:polfinalsymplecticfermion2}. Here $\mu_\la$ are the twist
operators defined in eq.~\eqref{eq:polsympltwist}. We have also used that
in going to this expectation value under the
rescaling~\eqref{eq:rescaling} we have to multiply the correlation
functions with an overall factor of $k$. This is because the correlation
function normalisation is relative to the correlator of $\bar\chi^1\chi^2$
or $c'_+c'_-$ in the $b'c'$ system in eq.~\eqref{eq:polfinal}. This simply
means that the dependence on $k$ disappears due to the normalisation in
eq.~\eqref{eq:newtwistfield} as is expected.

We want to compare this to the calculation of bulk twist correlators done
by Kausch in~\cite{Kausch:2000fu}. In that paper, of course, only twist
fields with identical twist in the holomorphic and anti-holomorphic sector
are treated so we take $\la_i=\bar\la_i$. Further, we have to remember
that the twist fields are only defined up to normalisation. To compare
with Kausch we use one of the equations~\eqref{eq:3pt5},~\eqref{eq:3pt6}
to fix the normalisation and can then compare to the second one. The
normalisation is fixed by defining
\begin{align}\label{eq:3pt7}
    \mu^L_{\la}\mu^R_{\la}=-\sqrt{\frac{\Ga(\la^*)}{\Ga(\la)}}\mu_{\la}.
\end{align}
Then we get
\begin{align}
    \expect{\mu_{\la_1}(z_1,\bar z_1)\mu_{\la_2}(z_2,\bar z_2)\mu_{\la_3}(z_3,\bar z_3)}_{\mathrm{SF}}&
    =\prod_{i}\sqrt{\frac{\Ga(\la_i)}{\Ga(\la_i^*)}}\prod_{i<j}\abs{z_i-z_j}^{2\la_i\la_j}\quad\textrm{for }\sum_i\la_i=1,\nonumber\\
    &=\prod_{i}\sqrt{\frac{\Ga(\la^*_i)}{\Ga(\la_i)}}\prod_{i<j}\abs{z_i-z_j}^{2\la^*_i\la^*_j}\quad\textrm{for }\sum_i\la_i=2,\label{eq:3pt8}
\end{align}
which is exactly as in~\cite{Kausch:2000fu}. We can also compare with the
two-point function which is easily calculated and also get a match here.
Note, however, that in~\cite{Kausch:2000fu} only ground state twist fields
with $0<\la<1$ are considered. Our results thus compare precisely in this
range, and are the analytic continuation of the twists $\la$ for the
results in~\cite{Kausch:2000fu}.

In the case where we allow the $e_i$ to be zero or an integer multiple of
$k$, we have to take into account the zero modes of the symplectic
fermions. This gives four different ground states in the symplectic model
- two fermionic and two bosonic, where the last two span a Jordan block
for $L_0$. The result is that we get logarithmic branch cuts in the
correlation functions. This can be seen from the \GL\ side where the $\Ga$
functions diverge when $\la$ becomes integer~\cite{Schomerus:2005bf}. Thus
we also get agreement from the two sides of the correspondence here.

\section{Branes in the symplectic fermions}

Now, having established the correspondence, we want to apply it. There are
two apparent applications. For point-like branes in the \GL\ WZNW model,
so far it could be argued that correlators containing only boundary fields
behave like untwisted symplectic fermions \cite{Creutzig:2008ek}, but it
was not possible to handle insertions of bulk fields. Now, we are in a
position to approach the problem of computing correlation functions
involving bulk and boundary fields. We will refrain from this problem for
now, but keep it in mind for future research. Instead, we reconsider the
study of boundary states. Recall that the group of outer automorphisms of
\GL\ is $\Z_2$. The branes corresponding to the trivial gluing
automorphism we call untwisted and their boundary states have been studied
in \cite{Creutzig:2007jy}. The non-trivial automorphism only admits one
volume-filling brane, which we call twisted. Its boundary state has been
studied, with quite some effort, in \cite{Creutzig:2008ek}. With the
\GL-symplectic fermion correspondence, we can easily reproduce these
results, but also compute spectra of strings stretching between an
untwisted and a twisted brane. This gives, finally, a complete discussion
of Cardy boundary states. It will turn out that the boundary states indeed
satisfy Cardy's condition, i.e. the amplitude is a true character.

As we have seen, the \GL\ WZNW model can equally well be understood in a
theory of scalars and symplectic fermions. Since boundary states with
symplectic fermions have not been discussed in completeness before, we
start by a quite general analysis of these. For earlier works on boundary
models of symplectic fermions
see~\cite{Bredthauer:2002ct,Bredthauer:2002xb,Kawai:2001ur,Gaberdiel:2006pp}.

\subsection{Boundary conditions}

We start our considerations by investigating possible boundary conditions.
The energy momentum tensors are
\begin{equation}
    T(z) \ = \  -\half\ep_{ab}\normord{\del\chi^a\del\chi^b}, \qquad \bar{T}(\bar{z}) \ = \ -\half\ep_{ab}\normord{\delbar\chi^a\delbar\chi^b}.
\end{equation}
They preserve the symplectic fermion symmetry and coincide along the boundary if
\begin{equation}\label{eq:bdyconditions}
    \begin{split}
        \del\chi \ = \ A\ \bar{\del}\chi \qquad\qquad {\rm for} \ z\ = \bar{z}\,,
\end{split}
\end{equation}
where $A=\bigl( \begin{smallmatrix}
a&b\\c&d
\end{smallmatrix} \bigr)$ is a matrix in SL(2) and for convenience we combined the two fermions in the vector $\chi = \Bigl( \begin{smallmatrix}
\chi^1\\ \chi^2
\end{smallmatrix} \Bigr)$.
In terms of Dirichlet and Neumann derivatives
($\del=\half\del_u-i\half\del_n$ and $\delbar=\half\del_u+i\half\del_n$)
the boundary conditions are
\begin{equation}
    \begin{split}
        -i\del_n \chi \ = \ \frac{A-1}{A+1}\ \del_u\chi\
\end{split}
\end{equation}
provided $1+A$ is invertible.
Then the action on the upper half-plane is
\begin{equation} \label{act}
    \begin{split}
    S \ = \ &-\frac{1}{4\pi} \int d^2z\ \del\chi^t\, J\,\bar{\del}\chi\ +\
     \frac{i}{8\pi} \int\limits_{z=\bar{z}} du \ \chi^t\, J\, \frac{A-1}{A+1}\, \del_u\chi\,,
     \end{split}
  \end{equation}
where the matrix $J$ is $J=\bigl( \begin{smallmatrix}
0&-1\\ 1&0
\end{smallmatrix} \bigr)$. The variation of this action vanishes provided the above boundary conditions hold as well as the bulk
equations of motion $\del\bar{\del}\chi^\pm=0$.
  If $1+A$ is not invertible it has characteristic polynomial $\lambda^2$,
i.e. if $1+A=0$ there are Dirichlet conditions in both directions while
otherwise there is one Dirichlet and one Neumann condition. Note that
these cases resemble the atypical branes in \GL\ \cite{Creutzig:2008ag}.

\subsection{The Ramond sector}

We first consider the Ramond sector, by which we mean the symplectic fermions without any twist fields,
or in the language of modes meaning only integer modes appear.
The explicit mode expansion is
\begin{equation}\label{eq:modeexpansionfermions}
     \chi^a(z,\bar{z}) \ = \ \xi^a+\chi^a_0\ln |z|^2 - \sum_{n\neq0}\frac{1}{n}\chi^a_nz^{-n}+\frac{1}{n}\bar{\chi}^a_n\bar{z}^{-n},
 \end{equation}
where the modes satisfy
\begin{equation} \label{eq:fermionrelations}
    \{ \chi^a_m,\chi^b_n\} \ = \ -m\ep^{ab}\,\delta_{m,-n} \ \ ,\ \  \{ \bar{\chi}^a_m,\bar{\chi}^b_n\} \ = \ -m \ep^{ab}\,\delta_{m,-n} \ \
    {\rm and}\ \  \{ \xi^a,\chi_0^b \} \ = \ \ep^{ab} \ \ .
\end{equation}
All other anti-commutators vanish. Note that for locality we have required $\chi_0^a=\bar\chi_0^a$.

In this section we construct the
boundary states in the Ramond sector, compute the amplitudes and construct
the corresponding open string model. We start the discussion of boundary
states by investigating Dirichlet conditions in the two fermionic
directions.

\subsubsection{Dirichlet conditions}

Let us first remind ourselves that if we have an extended chiral algebra
given by $W(z)$ and $\bar W(\bar z)$ we need an gluing automorphism,
$\Om$, for the boundary~\cite{Recknagel:1997sb}:
\begin{equation}\label{eq:gluingauto}
    W(z)=\Om(\bar W)(\bar z)\qquad\textrm{for } z=\bar z\,.
\end{equation}
This is as in equation~\eqref{eq:bdyconditions} for the gluing of the
currents. We now pass to closed strings via the world-sheet duality. The
gluing conditions then become the following Ishibashi conditions for the
boundary states $|\al\rrangle_{\Om}$ in the CFT on the full plane:
\begin{equation}\label{eq:ishibashigeneral}
    \left(W_n-(-1)^{h_W}\Om(\bar W_{-n})\right)|\al\rrangle_{\Om}\,,
\end{equation}
where $h_W$ is the conformal dimension of $W$.

Using~\eqref{eq:ishibashigeneral} we see that for the Dirichlet boundary
conditions ($A=-1$ in~\eqref{eq:bdyconditions}) the corresponding
Ishibashi states have to satisfy
\begin{equation}
    \begin{split}
    \left(\chi^a_n-\bar{\chi}^a_{-n}\right)|D\rrangle \ = \ 0\qquad\qquad\text{for}\ a=1,2\ ,
\end{split}
\end{equation}
note that there is no condition on $\chi^a_0$ because of the locality
constraint $\chi^a_0-\bar{\chi}^a_{0}=0$. The Ishibashi states are
explicitly constructed as
\begin{eqnarray}
    |D_0\rrangle \ &= \ \sqrt{2\pi}\exp\Bigl(\sum_{m>0}\frac{1}{m}
    \bigl(\chi^2_{-m}\bar{\chi}^1_{-m}-\chi^1_{-m}\bar{\chi}^2_{-m}\bigr)\Bigr)|0\rangle\,, \label{eq:D0}\\
    |D_\pm\rrangle \ &= \ \xi^\pm\ \exp\Bigl(\sum_{m>0}\frac{1}{m}\bigl(\chi^2_{-m}\bar{\chi}^1_{-m}-\chi^1_{-m}\bar{\chi}^2_{-m}\bigr)\Bigr)|0\rangle\,, \\
    |D_2\rrangle \ &= \ \frac{\xi^-\xi^+}{\sqrt{2\pi}}\ \exp\Bigl(\sum_{m>0}\frac{1}{m}
    \bigl(\chi^2_{-m}\bar{\chi}^1_{-m}-\chi^1_{-m}\bar{\chi}^2_{-m}\bigr)\Bigr)|0\rangle\,, \label{eq:D2}
\end{eqnarray}
where the ground state $|0\rangle$ is defined by $\chi^a_n|0\rangle=0$ for
$n\geq0$. The dual Ishibashi state is obtained by dualizing the modes
using (here $m>0$)
\begin{equation}
    {\chi^1_{-m}}^\dag \ = \ \chi^1_{m}\qquad\text{and}\qquad{\chi^2_{-m}}^\dag \ = \ -\chi^2_{m}\  .
\end{equation}
For the computation of amplitudes we need the Virasoro generators, they are
\begin{equation}
L_n\ = \ -\half\ep_{ab}\sum_{m} : \chi^a_{n-m}\chi^b_{m}:\
\end{equation}
and the central charge is $c=-2$.
Define $q=\exp 2\pi i \tau$ and $\tilde{q}=\exp( -2\pi i /\tau)$ as
usual, where $\tau$ takes values in the upper half plane. Then the
non-vanishing overlaps are
\begin{equation}\label{eq:Dirichletoverlap}
    \begin{split}
        \llangle D_0| q^{L_0^c+\frac{1}{12}}(-1)^{F^c}|D_2\rrangle \ &= \
        \llangle D_2| q^{L_0^c+\frac{1}{12}}(-1)^{F^c}|D_0\rrangle \ = \ \eta(\tau)^2, \\
        \llangle D_-| q^{L_0^c+\frac{1}{12}}(-1)^{F^c}|D_+\rrangle \ &= \
        -\llangle D_+| q^{L_0^c+\frac{1}{12}}(-1)^{F^c}|D_-\rrangle \ = \ \eta(\tau)^2, \\
                \llangle D_2| q^{L_0^c+\frac{1}{12}}(-1)^{F^c}|D_2\rrangle \ &= \ - i\tau\eta(\tau)^2\ = \ \eta(\tilde{\tau})^2\ , \\
    \end{split}
\end{equation}
where $L_0^c=L_0+\bar{L_0}$. Thus only $|D_2\rangle$ makes sense as a boundary state.

\subsubsection{Neumann conditions}

Next we would like to display the boundary state $|A\rangle$ for our
general boundary conditions \eqref{eq:bdyconditions}. It has to satisfy
the Ishibashi condition~\eqref{eq:ishibashigeneral}
\begin{equation}
    \begin{split}
    \chi^1_n+a\,\bar{\chi}^1_{-n}+b\,\bar{\chi}^2_{-n}|A\rrangle \ = \ 0\,, \\
    \chi^2_n+c\,\bar{\chi}^1_{-n}+d\,\bar{\chi}^2_{-n}|A\rrangle \ = \ 0\,, \\
\end{split}
\end{equation}
which are satisfied by
\begin{equation}
    |A\rrangle \ = \ \mathcal{N}\ \exp\Bigl(-\sum_{m>0}\frac{1}{m}
    \bigl(a\chi^2_{-m}\bar{\chi}^1_{-m}+b\chi^2_{-m}\bar{\chi}^2_{-m}-c\chi^1_{-m}\bar{\chi}^1_{-m}-d\chi^1_{-m}\bar{\chi}^2_{-m}\bigr)\Bigr)|0\rangle\,.
\end{equation}
The dual state is
\begin{equation}
     \llangle A| \ = \ \mathcal{N}\ \llangle0| \exp\Bigl(-\sum_{m>0}\frac{1}{m}
    \bigl(-a\chi^2_{m}\bar{\chi}^1_{m}+b\chi^2_{m}\bar{\chi}^2_{m}-c\chi^1_{m}\bar{\chi}^1_{m}+d\chi^1_{m}\bar{\chi}^2_{m}\bigr)\Bigr)\ .
\end{equation}
It will turn out that the normalization should be fixed to be
\begin{equation}\label{eq:normalization}
\mathcal{N}\ = \ \sqrt{2\pi}\, 2\, \sin\pi\mu\,,
\end{equation}
where we introduce $\mu$ via $\alpha=\exp 2\pi i\mu$ by $-\tr(A)=\alpha+\alpha^{-1}$.

Now it is straightforward to compute amplitudes between two boundary
states. Any non-zero amplitude requires the zero modes of $\chi^1$ and $\chi^2$ hence only the Dirichlet boundary state has
non-vanishing overlap with any Neumann state:
\begin{equation}\label{eq:RamondNeumannDirichletamplitude}
    \llangle A|\, q^{\frac{1}{2}L_0^c+\frac{1}{12}}\,(-1)^{F^c}\,|D_2\rrangle \ = \ \frac{\mathcal{N}}{\sqrt{2\pi}}q^{\frac{1}{12}}\prod_{m>0}(1-\alpha_{12} q^m)(1-\alpha_{12}^{-1} q^m)\,.
\end{equation}
Upon modular transformation this amplitude is the spectrum of an open
string stretching between two branes with respectively Neumann boundary
conditions given by $A$ and Dirichlet conditions. Using the formulas
provided in the appendix equation
\eqref{eq:RamondNeumannDirichletamplitude} becomes
\begin{equation}
    \begin{split}
        \frac{\mathcal{N}}{\sqrt{2\pi}}\,q^{\frac{1}{12}}\prod_{m>0}(1-\alpha q^m)(1-\alpha^{-1} q^m) \
        = \ \tilde{q}^{\frac{1}{2}(\mu-\frac{1}{2})^2-\frac{1}{24}}\,\prod_{n=0}^\infty
        \bigl(1-\tilde{q}^{n+1-\mu}\bigr)\bigl(1-\tilde{q}^{n+\mu}\bigr)\, .\\
    \end{split}
\end{equation}
Now, we construct the boundary theory of a string stretching between these
two branes and check that its spectrum is indeed given by the amplitude we
just computed, we follow~\cite{Creutzig:2006wk}. Therefore consider the upper half plane, and demand
boundary condition $A$ for the negative real line, i.e.
\begin{equation}
    \del\chi \ = \ A\,\bar{\del}\chi    \qquad{\rm for}\ z\ = \bar{z} \qquad{\rm and}\ z+\bar{z}\ <\ 0\ ;
\end{equation}
and Dirichlet conditions for the positive real axis
\begin{equation}
    \del_u\chi \ = \ 0  \qquad{\rm for}\ z\ = \bar{z} \qquad{\rm and}\ z+\bar{z}\ >\ 0\ .
\end{equation}
Then the fields have the following SL(2) monodromy (counterclockwise)
\begin{equation}
    \del\chi(ze^{2\pi i}) \ = \ -A\del\chi(z)\,,
\end{equation}
and similar for the bared quantities. Denote by $S$ the matrix that
diagonalizes the monodromy, i.e. $S(-A)S^{-1}$ is diagonal. We denote the
eigenvalues by $\alpha^{\pm1}$. Further, call the eigenvectors
$\del{\chi}^\pm$, they then have the usual mode expansion \cite{Kausch:2000fu}
\begin{equation}
    {\chi}^\pm(z) \ = \ \sum_{n\in\Z}\frac{1}{n\pm\mu}{\chi}^\pm_{n\pm\mu}z^{-(n\pm\mu)}\, .
 \end{equation}
The original fields are then explicitly
\begin{equation}
    \begin{pmatrix}\chi^1\\ \chi^2\\\end{pmatrix} \ = \ S^{-1} \begin{pmatrix}{\chi}^+\\ {\chi}^-\\\end{pmatrix}\, .
\end{equation}
Their partition function is
\begin{equation}
    \tr (\, q^{L_0-\frac{c}{24}}(-1)^F\, ) \ = \
    q^{\frac{1}{2}(\mu-\frac{1}{2})^2-\frac{1}{24}}\prod_{n=0}^\infty
        \bigl(1-q^{n+1-\mu}\bigr)\bigl(1-q^{n+\mu}\bigr)\,.
\end{equation}
The computation has been done similarly by Kausch \cite{Kausch:2000fu}. We see
that the result fits with~\eqref{eq:RamondNeumannDirichletamplitude} and
the Cardy condition is fulfilled. Thus, we nicely established our boundary
state and the open string theory it describes.

If we want to investigate amplitudes involving Neumann boundary states on
both ends, we learnt \cite{Creutzig:2006wk} that it is necessary to insert
additional zero modes in order to obtain a non-vanishing amplitude. Also
introduce $\alpha_{12}$ via
$\tr(A_1A_2^{-1})=\alpha_{12}+\alpha_{12}^{-1}$ then we get
\begin{equation}\label{eq:Ramondamplitude}
\begin{split}
    \llangle A_1|\, \chi^2\chi^1\, q^{\frac{1}{2}L_0^c+\frac{1}{12}}\,(-1)^{F^c}\,|A_2\rrangle \ &= \
\mathcal{N}_1\mathcal{N}_2\,q^{\frac{1}{12}}\prod_{m>0}(1-\alpha_{12} q^m)(1-\alpha_{12}^{-1} q^m)\\
        &= \ \mathcal{N}_{12}\,\tilde{q}^{\frac{1}{2}(\mu_{12}-\frac{1}{2})^2-\frac{1}{24}}\,\prod_{n=0}^\infty
        \bigl(1-\tilde{q}^{n+1-\mu_{12}}\bigr)\bigl(1-\tilde{q}^{n+\mu_{12}}\bigr)\,,
    \end{split}
\end{equation}
where
\begin{equation}
       \mathcal{N}_{12}\ = \ 4\pi\, \frac{\sin\pi\mu_1\, \sin\pi\mu_2}{\sin\pi\mu_{12}}\, .
\end{equation}
The open string theory is constructed almost exactly as above and again
resembles \cite{Creutzig:2006wk}. We demand boundary condition $A_1$ for
the negative real line and $A_2$ for the positive one,
\begin{equation}
    \del\chi \ = \ \left\{ \begin{array}{ll}
         A_1\,\bar{\del}\chi & \qquad{\rm if}\ z\ = \bar{z} \qquad{\rm and}\ z+\bar{z}\ <\ 0\\
         A_2\,\bar{\del}\chi & \qquad{\rm if}\ z\ = \bar{z} \qquad{\rm and}\ z+\bar{z}\ >\ 0\ .\end{array} \right.
\end{equation}
The fields have the following SL(2) monodromy
\begin{equation}
    \del\chi(ze^{2\pi i}) \ = \ A_1A_2^{-1}\,\del\chi(z)\, .
\end{equation}
Let $S$ diagonalize the monodromy, then its eigenvalues are
$\alpha_{12}^{\pm1}$ and we call the eigenvectors again $\del{\chi}^\pm$.
They have the mode expansion
\begin{equation}
    {\chi}^\pm(z) \ = \ \sqrt{\mathcal{N}_{12}}\,{\xi}^\pm+
    \sum_{n\in\Z}\frac{1}{n\pm\mu_{12}}{\chi}^\pm_{n\pm\mu_{12}}z^{-(n\pm\mu_{12})}\,,
 \end{equation}
note the extra zero mode, since the monodromy does only concern
derivatives. Its partition function with appropriate insertion is
\begin{equation}
    \tr (\chi^2\chi^1q^{L_0-\frac{c}{24}}(-1)^F ) \ = \ \mathcal{N}_{12}\,
    q^{\frac{1}{2}(\mu_{12}-\frac{1}{2})^2-\frac{1}{24}}\prod_{n=0}^\infty
        \bigl(1-q^{n+1-\mu_{12}}\bigr)\bigl(1-q^{n+\mu_{12}}\bigr)\,,
\end{equation}
and coincides with \eqref{eq:Ramondamplitude} as desired.

\subsection{The Neveu-Schwarz sector}

In this section we study the boundary states in the Neveu-Schwarz sector.
The states have to satisfy the usual Ishibashi condition
\begin{equation}
    \begin{split}
    \chi^1_n+a\bar{\chi}^1_{-n}+b\bar{\chi}^2_{-n}|A\rrangle_{NS} \ = \ 0 \,,\\
    \chi^2_n+c\bar{\chi}^1_{-n}+d\bar{\chi}^2_{-n}|A\rrangle_{NS} \ = \ 0 \,,
\end{split}
\end{equation}
where the modes are half-integer, i.e. $n$ in $\Z+1/2$.
The conditions are satisfied by
\begin{equation}
    |A\rrangle \ = \ \exp\Bigl(-\sum_{\substack{m>0 \\ m \, \in\, \Z+1/2}}\frac{1}{m}
    \bigl(a\chi^2_{-m}\bar{\chi}^1_{-m}+b\chi^2_{-m}\bar{\chi}^2_{-m}-c\chi^1_{-m}\bar{\chi}^1_{-m}-d\chi^1_{-m}\bar{\chi}^2_{-m}\bigr)\Bigr)|0\rangle\,.
\end{equation}
We introduce $\alpha_{12}$ as before, that is
$\tr(A_1A_2^{-1})=\alpha_{12}+\alpha_{12}^{-1}$, and get
\begin{equation}
    \begin{split}
    {}_{NS}\llangle A_1| q^{L_0^c+\frac{1}{12}}(-1)^{F_c}|A_2\rrangle_{NS} \ &= \
    q^{-\frac{1}{24}}\prod_{\substack{m>0 \\ m \, \in\, \Z+1/2}}(1-\alpha_{12} q^m)(1-\alpha_{12}^{-1} q^m)\\
    &= \    \tilde{q}^{\frac{1}{2}(\mu-\frac{1}{2})^2-\frac{1}{24}}\prod_{n>0}(1+\tilde{q}^{n-\mu})
        (1+\tilde{q}^{n-\mu^*})\,,
    \end{split}
\end{equation}
where $\alpha_{12}=e^{2\pi i\mu}$. This is the spectrum of an open string constructed similarly as before,
and in addition demanding antisymmetric boundary conditions in the time-direction.

\subsection{The twisted sectors}

Given any twisted sector we can diagonalize it and thus we can restrict to
twists that are diagonal. Call the ground state of the sector for
$\mu_\lambda$ on which $\chi^a$ has twists
\begin{equation}
    \chi^1 \ \longrightarrow \ e^{-2\pi i\lambda}\chi^1\qquad\text{and}\qquad\chi^2 \ \longrightarrow \ e^{2\pi i\lambda}\chi^2\,.
\end{equation}
Then the mode expansions of the fields in these sectors are
\begin{equation}\label{eq:twistedmodeexpansion}
\begin{split}
    \del\chi^1(z) \ = \ -\sum_{n\in\Z}\chi^1_{n+\lambda}z^{-(n+\lambda)-1}\qquad{\rm and}\qquad
    \bar{\del}\bar{\chi}^1(\bar{z}) \ = \ -\sum_{n\in\Z}\bar{\chi}^1_{n-\lambda}\bar{z}^{-(n-\lambda)-1}\\
    \del\chi^2(z) \ = \ -\sum_{n\in\Z}\chi^2_{n-\lambda}z^{-(n-\lambda)-1}\qquad{\rm and}\qquad
    \bar{\del}\bar{\chi}^2(\bar{z}) \ = \ -\sum_{n\in\Z}\bar{\chi}^2_{n+\lambda}\bar{z}^{-(n+\lambda)-1}\,.
\end{split}
\end{equation}
Whenever $\lambda\neq1/2$ the boundary conditions are parameterized by
just one parameter $\alpha$ according to the boundary conditions
\begin{equation}
    \begin{split}
        \del \chi^1 \ = \ \alpha\bar{\del}\chi^1\qquad{\rm and}\qquad
    \del \chi^2 \ = \ \alpha^{-1}\bar{\del}\chi^2 \ .
\end{split}
\end{equation}
Only to these conditions there exist twisted Ishibashi states. The
boundary state has to satisfy the usual Ishibashi condition
\begin{equation}
    \begin{split}
        \chi^1_{n+\lambda}+\alpha\bar{\chi}^1_{-n-\lambda}|\alpha\rrangle_{\lambda} \ = \ 0\,, \\
        \chi^2_{n-\lambda}+\alpha^{-1}\bar{\chi}^2_{-n+\lambda}|\alpha\rrangle_{\lambda} \ = \ 0\,,
\end{split}
\end{equation}
and these are solved by ($\lambda^*=1-\lambda$)
\begin{equation}\label{eq:twistedIshibashistate}
    |\alpha\rrangle_{\lambda} \ = \ \mathcal{N}\exp\Bigl(-\sum_{m>0}\frac{\alpha}{m-\lambda^*}\chi^2_{-m+\lambda^*}\bar{\chi}^1_{-m+\lambda^*}
    -\frac{\alpha^{-1}}{m-\lambda}\chi^1_{-m+\lambda}\bar{\chi}^2_{-m+\lambda}\Bigr)\mu_{\lambda} \ .
\end{equation}
where we fix the normalization to be $\mathcal{N}=e^{-2\pi i(\lambda-1/2)(\mu-1/4)}$ and $\alpha=e^{2\pi i \mu}$.
The dual boundary state is
\begin{equation}
    {}_{\lambda}\llangle\alpha | \ = \ \mathcal{\bar N}\mu_{\lambda}^\dag\exp\Bigl(\sum_{m>0}\frac{\alpha}{m-\lambda}\chi^2_{m-\lambda}\bar{\chi}^1_{m-\lambda}
    -\frac{\alpha^{-1}}{m-\lambda^*}\chi^1_{m-\lambda^*}\bar{\chi}^2_{m-\lambda^*}\Bigr)\,.
\end{equation}
Now we are prepared to compute the amplitudes (note that the conformal
dimension of the twist state is $h_\lambda=-\lambda\lambda^*/2$ and we use
the shorthand $\alpha_1\alpha_2^{-1}=e^{2\pi i\mu}$)
\begin{equation}
    \begin{split}
        {}_\lambda\langle\alpha_1| q^{L_0^c+\frac{1}{12}}(-1)^{F_c}|\alpha_2\rangle_\lambda \ &= \ e^{-2\pi i(\lambda-\frac{1}{2})(\mu-\frac{1}{2})}
        q^{\frac{1}{2}(\lambda-\frac{1}{2})^2-\frac{1}{24}}\prod_{n>0}(1-\alpha_1\alpha_2^{-1}q^{n-\lambda^*})
        (1-\alpha_2\alpha_1^{-1}q^{n-\lambda}) \\
        &= \ \tilde{q}^{\frac{1}{2}(\mu-\frac{1}{2})^2}
        \theta\Bigl(\tilde{\tau}(\frac{1}{2}-\mu)-(\lambda-\frac{1}{2}),\tilde{\tau}\Bigr)/\eta(\tilde{\tau})\\
        &= \ \tilde{q}^{\frac{1}{2}(\mu-\frac{1}{2})^2-\frac{1}{24}}\prod_{n>0}(1-u^{-1}\tilde{q}^{n-\mu})
        (1-u\tilde{q}^{n-\mu^*})\,,
    \end{split}
\end{equation}
where $u=e^{2\pi i\lambda}$. This is the character of a boundary theory
twisted by $\mu_{12}$ in an orbifold by an abelian subgroup $\mathcal{G}$ of SL(2),
which is generated by $u$, see \cite{Kausch:2000fu} for a
detailed discussion.

\section{Branes in the GL(1$|$1) WZNW model}

We are finally in a position to apply the symplectic fermion \GL\
correspondence to boundary states in \GL. \GL\ possesses two non-trivial
gluing conditions. One condition, which we call untwisted since it is
given by a trivial gluing automorphism, has a two-parameter family of
branes corresponding to super conjugacy classes. They have been discussed
in detail in \cite{Creutzig:2007jy}. The other gluing condition, which we
call twisted, consists of one volume filling brane. Its boundary state has
been discussed in \cite{Creutzig:2008ek} as a rather complicated
perturbative expansion from a free scalar times a symplectic fermion
boundary state. In~\cite{Creutzig:2008ek} it was shown that for a
particular amplitude this expansion did not contribute and computations
reduced to computations in the decoupled free scalar free fermion model.
This was already a first hint of the correspondence now found.

We can now compute amplitudes between boundary states of different gluing
conditions and construct the corresponding open string theory explicitly.
There is another puzzle we can unreveal and that is the role of atypical
Ishibashi states and their $\log q$ dependent overlaps. While Ishibashi
states corresponding to typical representations have true characters as
overlap, atypicals might have a $\log q$ prefactor, as seen in
\eqref{eq:Dirichletoverlap} and e.g.~\cite{Kogan:2000fa}. We will see that in the \GL\ story, this
$\log q$ dependence arises by a limiting procedure from characters of
typical representations. The understanding of the atypical Ishibashi
states in our context is important for amplitudes involving two branes
with different gluing conditions.

\subsection{Untwisted Branes}

Let us recall the analysis performed in \cite{Creutzig:2007jy}. Untwisted
branes correspond to the gluing conditions
\begin{equation}
    J(z)\ = \bar{J}(\bar{z})\qquad{\rm for}\ z \ = \ \bar{z}\,.
\end{equation}
Insertion of the explicit formulas \eqref{eq:finalcurrent}
for the currents into above gluing
conditions gives Dirichlet conditions for the bosonic currents
\begin{equation} \label{eq:BUG}
    \begin{split}
        &\del_u Y \ = \ 0 \qquad , \qquad \del_u Z\  = \ 0 \qquad ,
         \qquad \text{for} \ z\ =\ \bar{z} \, ,  \\
        \end{split}
\end{equation}
While the fermionic ones generically satisfy Neumann conditions
\begin{equation} \label{eq:FUG}
    \begin{split}
        e^{Y^L_0}\del\chi^1\ = \ -e^{-Y^R_0}\bar\del\chi^1\qquad\text{and}\qquad e^{-Y^L_0}\del\chi^2\ = \ -e^{Y^R_0}\bar\del\chi^2\, .
    \end{split}
\end{equation}
Thus one parameterizes the branes by their positions labeled by
$(y_0,z_0)$. But whenever\footnote{In this section we allow for imaginary
brane positions exactly as done in \cite{Creutzig:2007jy}.}
$Y_0^L+Y_0^R=iy_0 = 2\pi is, s \in \mathbb{Z}$ we obtain Dirichlet
boundary conditions in all directions, bosonic and fermionic ones,
\begin{equation}
    \del_u Y\ =\ \del_u Z\ =\ \chi^a\ =\ 0 \ \ \text{for} \ z\, =\, \bar{z}\,.
\end{equation}
These branes will be called non-generic (untwisted) branes in the following.
\smallskip

Let us shortly describe the main results of the minisuperspace analysis
performed in \cite{Creutzig:2007jy}. The minisuperspace or particle limit
describes the behaviour of full field theory quantities in the large level
limit, $k\rightarrow\infty$. In this limit the zero modes of the fields
dominate and thus fields are interpreted as functions on the supergroup,
and the action of the currents is mimicked by the right and left invariant
vector fields $R_X$ and $L_X$. We are interested in semiclassical analogua
of Ishibashi states. Minisuperspace Ishibashi states are those states
invariant under the adjoint action $\ad_X=R_X+L_X$ since the gluing
automorphism is the identity. What is interesting for our further
consideration is that there exist two kinds of atypical Ishibashi states.
One of them has vanishing overlap with itself and is thus associated with
\eqref{eq:D0} in our correspondence to symplectic fermions, while the
other one is obtained from the first kind by the action of the fermionic
functions associated to the fermionic fields $c_\pm$, and thus it should
be identified with \eqref{eq:D2}. Further, in the minisuperspace limit
boundary states become distributions concentrated on the super conjugacy
class they correspond to, and this distribution can be expressed in terms
of minisuperspace Ishibashi states. It turns out that the first kind of
Ishibashi state contributes to the generic boundary states while the
second kind to the non-generic one. For more details, we refer the reader
to \cite{Creutzig:2007jy}, but we will also illustrate the lift of this
minisuperspace story to the full field theory in the following
subsections.

\subsubsection{Ishibashi states}

We now construct the Ishibashi states using our symplectic fermion
correspondence. Recall that the currents take the
form~\eqref{eq:finalcurrent}
\begin{align}
        J^E &= -k\del Y, & J^N &= -\del Z, &  J^- &= \sqrt{k}e^{Y^L}\del\chi^1, &   J^+ &= -\sqrt{k}e^{-Y^L}\del\chi^2, \\
        \bar{J}^E &= k\bar{\del} Y, & \bar{J}^N &=  \delbar Z, &
        \bar{J}^- &= -\sqrt{k}e^{-Y^R}\delbar\chi^1, &  \bar{J}^+ &= \sqrt{k}e^{Y^R}\delbar\chi^2.
\end{align}
Further, the fermions have mode expansion as in equation
\eqref{eq:modeexpansionfermions} and relations \eqref{eq:fermionrelations}
(or the twisted versions thereof) while the two scalars have expansion
\begin{equation}
    \begin{split}
        Y^L(z) \ &= \ Y_0^L + p_Y^L\, \ln z - \sum_{n\neq0}\, \frac{1}{n}\,  Y^L_n\, z^{-n},  \\
        Y^R(z) \ &= \ Y_0^R + p_Y^R\, \ln \zbar - \sum_{n\neq0}\, \frac{1}{n}\,  Y^R_n\, \zbar^{-n},  \\
        Z^L(z) \ &= \ Z_0^L + p_Z^L\, \ln z - \sum_{n\neq0}\, \frac{1}{n}\,  Z^L_n\, z^{-n},  \\
        Z^R(z) \ &= \ Z_0^R + p_Z^R\, \ln \zbar - \sum_{n\neq0}\, \frac{1}{n}\,  Z^R_n\, \zbar^{-n},
    \end{split}
\end{equation}
and relations
\begin{equation}
    [Y_n^{L,R},Z_m^{L,R}]\ = \ -m\delta_{n,-m}\qquad{\rm and}\qquad [Z_0^{L,R},p_Y^{L,R}]\ = \ [Y_0^{L,R},p_Z^{L,R}]\ = \ -1\, .
\end{equation}
To ensure locality we have $p_Y^L=p_Y^R$ and also $Z_0^L=Z_0^R$ for the
conjugate modes. However, we will not demand $p_Z^L=p_Z^R$ and
correspondingly not $Y_0^L=Y_0^R$ since $Z$ has an additive twist around
our winding states~\eqref{eq:finalvertexoperator}.

The energy momentum tensor is
\begin{equation}
    T(z)\ = \ \del Y\del Z -\half\epsilon_{ab}\del\chi^a\del\chi^b\qquad{\rm and}\qquad
    \bar{T}(\zbar)\ = \ \delbar Y\delbar Z -\half\epsilon_{ab}\delbar\chi^a\delbar\chi^b\, ,
\end{equation}
and thus the Virasoro modes are
\begin{equation}
    \begin{split}
    L_n \ = \ &-\sum_{m\, \in\,\Z}\, \normord{\chi^1_{n-m}\chi^2_m} +  \sum_{m\,\neq\,0,n} \normord{Y^L_{n-m}Z^L_m}+\\
            &+\sum_{m\,\neq\,0}(\normord{p_Y^L Z^L_m}+\normord{p_Z^L Y^L_m})+\de_{n,0}\,p^L_Yp^L_Z\,,\\
    \bar{L}_n \ = \ &-\sum_{m\, \in\,\Z}\, \normord{\bar\chi^1_{n-m}\bar\chi^2_m} +  \sum_{m\,\neq\,0,n} \normord{Y^R_{n-m}Z^R_m}+\\
            &+\sum_{m\,\neq\,0}(\normord{p_Y^R Z^R_m}+\normord{p_Z^R Y^R_m})+\de_{n,0}\,p^R_Yp^R_Z \,.\\
\end{split}
\end{equation}
We also need the zero modes of the currents corresponding to the Cartan generators $J^E$ and $J^N$:
\begin{equation}
    E_0 \ = \ -kp_Y^L,\qquad\bar{E}_0 \ = \ kp_Y^R,\qquad N_0 \ = \ -p_Z^L ,\qquad \bar{N}_0 \ = \ p_Z^R\, .
\end{equation}
Let us now consider the Ishibashi states. We start by spelling out the
Ishibashi conditions for the untwisted case. As noted above, the gluing
condition $J=\bar{J}$ means that the bosonic fields simply satisfy
Dirichlet conditions
\begin{equation}
    \del_u Y \ = \ \del_u Z \ = \ 0\, .
\end{equation}
Using these Dirichlet conditions for the field $Y=Y^L+Y^R$ the fermionic ones can be written as follows
\begin{equation}
 e^{Y_0^L}\del\chi^1 \ = \ -e^{-Y_0^R}\delbar\chi^1
    \qquad{\rm and}\qquad e^{-Y_0^L}\del\chi^2 \ = \ -e^{Y_0^R}\delbar\chi^2\, .
\end{equation}
Then correspondingly the Ishibashi conditions for the bosonic fields are
\begin{equation}
\begin{split}
    \left(Y^L_n-Y^R_{-n}\right)\, |\, I \,\rrangle \ &= \ \left(Z^L_n-Z^R_{-n}\right)\, |\, I \,\rrangle \ = \ 0\qquad\ n\neq0\\
    \left(p^L_Z-p^R_{Z}\right)\, |\, I \,\rrangle \ &= \left(p^L_Y-p^R_{Y}\right)\, |\, I \,\rrangle \ = \ 0\, ,\\
    \end{split}
\end{equation}
note that there is no conditions on the zero modes $Y^L_0$ and $Y^R_0$.
Further, the conditions for the fermionic ones are
\begin{equation}
    \big(e^{Y_0^L}\chi^1_n-e^{-Y_0^R}\bar\chi^1_{-n}\big)\, |\, I \,\rrangle \ = \
    \big(e^{-Y_0^L}\chi^2_n-e^{Y_0^R}\bar\chi^2_{-n}\big)\, |\, I \,\rrangle \ = \ 0 \, .
\end{equation}
The Ishibashi states clearly factorize into a bosonic and a fermionic part
and are easily constructed as follows. The typical primary of \GL,
$\<e,n\>_R$, is the representation with ground state
$|n,\mu_\lambda\rangle$ where $\lambda=e/k$ satisfying
\begin{equation}
    \begin{split}
    p_Z^L|n,\mu_\lambda\rangle \ &= \ p_Z^R|n,\mu_\lambda\rangle \ = \ n|n,\mu_\lambda\rangle\,, \\
    p_Y^L|n,\mu_\lambda\rangle \ &= \ p_Y^R|n,\mu_\lambda\rangle \ = \ \lambda |n,\mu_\lambda\rangle \, .
\end{split}
\end{equation}
Further, recall that the fermions have the mode expansion in the presence
of the ground state $\mu_\lambda$ \eqref{eq:twistedmodeexpansion}
\begin{equation}
    \begin{split}
        \chi^1(z,\bar{z})\ &= \ \sum_{n\,\in\,\Z+\lambda}\,\frac{1}{n}\,\chi^1_{n}\,z^{-n}\,+\,
                                \sum_{n\,\in\,\Z+\lambda^*}\,\frac{1}{n}\,\bar{\chi}^1_{n}\,\zbar^{-n}\,,\\
        \chi^2(z,\bar{z})\ &= \ \sum_{n\,\in\,\Z+\lambda^*}\,\frac{1}{n}\,\chi^2_{n}\,z^{-n}\,+\,
                                \sum_{n\,\in\,\Z+\lambda}\,\frac{1}{n}\,\bar{\chi}^2_{n}\,\zbar^{-n}\,,\\
    \end{split}
\end{equation}
where $\lambda^*=1-\lambda$.
Then the bosonic Ishibashi state is
\begin{equation}
    |n,e\rrangle_B \ = \  \exp\Bigl(\sum_{m>0}\frac{1}{m}\bigl(Y^L_{-m}Z^R_{-m}+Z^L_{-m}Y^R_{-m}\bigr)\Bigr)|n,\mu_\lambda\rangle_B\,, \\
\end{equation}
and the fermionic one is computed as \eqref{eq:twistedIshibashistate}
\begin{equation}\label{eq:twistedGLIshibashistate}
    |n,e\rrangle_F \ = \ \exp\Bigl(-\sum_{m\,>\,0}
    \frac{e^{Y_0^L+Y_0^R}}{m-\lambda}\chi^1_{-m+\lambda}\bar{\chi}^2_{-m+\lambda}
    -\frac{e^{-Y_0^L-Y_0^R}}{m-\lambda^*}\chi^2_{-m+\lambda^*}\bar{\chi}^1_{-m+\lambda^*}\Bigr)|n,\mu_\lambda\rangle_F \ .
\end{equation}
and the Ishibashi state is then the product of the two. The following
simple computations are crucial
\begin{equation}
    \begin{split}
        q^{L_0}e^{\pm Y^L_0} \ &= \ e^{\pm Y_0^L}q^{L_0\mp \frac{E_0}{k}} \qquad , \qquad
        Z^{N_0}e^{\pm Y^L_0} \ = \ e^{\pm Y^L_0}Z^{N_0\mp1}\,, \\
        q^{\bar{L}_0}e^{\pm Y^R_0} \ &= \ e^{\pm Y_0^R}q^{\bar{L}_0\pm \frac{\bar{E}_0}{k}} \qquad , \qquad
        Z^{\bar{N}_0}e^{\pm Y^R_0} \ = \ e^{\pm Y^R_0}Z^{\bar{N}_0\pm1}\,, \\
    \end{split}
\end{equation}
Introduce $L_0^c=\frac{1}{2}(L_0+\bar{L}_0)$ and
$N_0^c=\frac{1}{2}(N_0-\bar{N}_0)$ as usual. Then we get the fermionic contribution
of the overlap, that is
\begin{equation}
    _F\llangle n,e|q^{L_0^c+\frac{1}{12}}z^{N_0^c}(-1)^{F^c} |n,e\rrangle_F \ = \
    z^{n}(1-z^{-1})q^{\frac{1}{2}(\lambda-\frac{1}{2})^2-\frac{1}{24}}\prod_{n>0}(1-z^{-1}q^n)(1-zq^n)\,,
\end{equation}
and the bosonic
 \begin{equation}
    _B\llangle n,e|q^{L_0^c-\frac{1}{12}}z^{N_0^c}(-1)^{F^c} |n,e\rrangle_B \ = \
    -\frac{q^{n\lambda}}{\eta(\tau)^2}\,,
\end{equation}
where we normalized the dual state such that we get the minus sign. Then
in total, we arrive at
\begin{equation}
    \begin{split}
        \llangle n,e|q^{L_0^c}z^{N_0^c}(-1)^{F^c}|n,e\rrangle \ &= \
    z^{n-1}(1-z)\frac{q^{n\lambda+\frac{1}{2}(\lambda-\frac{1}{2})^2-\frac{1}{24}}}{\eta(\tau)^2}\prod_{n>0}(1-z^{-1}q^n)(1-zq^n) \\
    &= \ \hat\chi_{<e,n>}(z,\tau)\,.
    \end{split}
\end{equation}
So far we assumed $0<\lambda<1$, whenever $\lambda$ becomes zero our
Dirichlet symplectic fermion boundary states come into the game. There are
four of them. Denote by $|n,0\rangle$ the ground state with $N_0$
eigenvalue $n$, i.e.
\begin{align}
    N_0|n,0\rangle&=n|n,0\rangle,\qquad  E_0|n,0\rangle=0\,,\nonumber \\
    Y_m|n,0\rangle&=Z_m|n,0\rangle=\chi^a_m|n,0\rangle=\chi^a_0|n,0\rangle=0,\quad\textrm{for }m>0.
\end{align}
Then the Ishibashi states are
\begin{equation}
    \begin{split}
        |n_0\rrangle \ &= \
    \exp\Bigl(\sum_{m>0}\frac{1}{m}\bigl(Y^L_{-m}Z^R_{-m}+Z^L_{-m}Y^R_{-m}-
    e^{Y^L_0+Y_0^R}\chi^1_{-m}\bar{\chi}^2_{-m}+e^{-Y^L_0-Y^R_0}\chi^2_{-m}\bar{\chi}^1_{-m}\bigr)\Bigr)|n,0\rangle\,, \\
    |n_\pm\rrangle \ &= \ \xi^\pm|n_0\rrangle\,, \\
    |n\rrangle \ &= \ \xi^-\xi^+|n_0\rrangle\,,
\end{split}
\end{equation}
and we arrive at the following amplitudes
\begin{equation}\label{eq:ishiamplin}
    \begin{split}
        \llangle n_0|q^{L^c_0}z^{N^c_0}(-1)^{F^c}|n\rrangle \ &= \ \chi_0(\mu,\tau), \\
                \llangle n|q^{L^c_0}z^{N^c_0}(-1)^{F^c}|n_0\rrangle \ &= \ -\chi_0(\mu,\tau), \\
                \llangle n_\pm|q^{L^c_0}z^{N^c_0}(-1)^{F^c}|n_\mp\rrangle \ &= \ -\chi_0(\mu,\tau), \\
                \llangle n|q^{L^c_0}z^{N^c_0}(-1)^{F^c}|n\rrangle \ &= \ -2\pi i\tau\chi_0(\mu,\tau), \\
    \end{split}
\end{equation}
where
\begin{equation}
    \chi_0(\mu,\tau)=z^{n-1}q^{\frac{1}{12}}\prod_{n>0}(1-z^{-1}q^n)(1-zq^n)/\eta(\tau)^2.
\end{equation}
All other amplitudes vanish unless zero modes are inserted.

Let us now consider twist states $\mu_{\tilde\lambda}$ where
$\tilde\lambda\not\in\ ]0,1[\,$. We already saw in the second section that
such states are simply descendants of $\mu_\lambda$ where $\tilde\lambda
=\lambda+m$ for some integer $m$ and $\la\in\,]0,1[$. The state
$|n,\mu_{\tilde\lambda}\rangle$ satisfies the following conditions
\begin{equation}
      N_0|n,\mu_{\tilde\lambda}\rangle\ = \ n|n,\mu_{\tilde\lambda}\rangle\qquad \text{and} \qquad E_0|n,\mu_{\tilde\lambda}\rangle \ = \ k(\lambda+m)|n,\mu_{\tilde\lambda}\rangle\,.
\end{equation}
The Ishibashi state $|e,n\rrangle$ (with $e/k=\tilde\lambda=\lambda+m$) in this representation is obtained from the previously constructed ones as
\begin{equation}
        |e,n\rrangle \ = \ e^{m(Z_0^L-Z_0^R)}e^{m(Y_0^L+Y_0^R)}|e-mk,n\rrangle\, .
\end{equation}
The amplitude is computed using
\begin{equation}
        q^{L_0^c}e^{m(Z_0^L-Z_0^R)}\ =\ e^{m(Z_0^L-Z_0^R)}q^{L_0^c-mN_0^c}\,,
\end{equation}
and the spectral flow formulas provided in appendix~\ref{app:representation}
\begin{equation}
    \begin{split}
        \llangle n,e|q^{L_0^c}z^{N_0^c}(-1)^{F^c}|n,e\rrangle \ &= \ \hat\chi_{<e-mk,n+m>}(z-m\tau,\tau) \ = \ (-1)^m\hat\chi_{<e,n>}(z,\tau)\, .
    \end{split}
\end{equation}
A similar construction holds also for the atypical part.

\subsection{The untwisted boundary states}

Untwisted boundary states were studied in detail in
\cite{Creutzig:2007jy}, here we recall the states and their properties. Atypical Ishibashi
states contribute to amplitudes only by a set of measure zero, therefore
their role has not been fully investigated previously. Here
we will fill this gap. As we will see in a moment boundary states are
represented by an integral of Ishibashi states, hence any amplitude is
given by an integral of \GL\ characters. We fix the role of the Ishibashi
states by requiring that any integrand of any amplitude is a smooth
function. Let us be more precise.

The minisuperspace analysis \cite{Creutzig:2007jy} already suggests that
the Ishibashi states $|n_0\rrangle$ contribute to generic branes, while
the states $|n\rrangle$ contribute to non-generic branes. We will see that
this is correct. The boundary state corresponding to a generic brane
localized at $(z_0,y_0)$ with $y_0 \neq 2\pi s$ is
\begin{equation}\label{eq:bst1}
    \begin{split}
    |z_0,y_0\rangle\ =\ &\sqrt{\frac{2i}{k}}\int_{\substack{e\neq mk\\ m\, \in \, \Z}} de dn\
     \exp\bigl(i(n-1/2)y_0+iez_0\bigr)\ \sin^{1/2}(\pi e/k)\ |e,n\rrangle \ - \\
     &\frac{\sqrt{2\pi i}}{k}\sum_{m\,\in\,\Z}\int dn\
     \exp\bigl(i(n-1/2)y_0+imkz_0\bigr)\ |n_0\rrangle^{(m)},
     \end{split}
\end{equation}
where the superscript $m$ denotes the $\ga_m$ spectral flowed state. In order
to check the consistency of our proposal for the boundary states with
world-sheet duality, we compute the spectrum between a pair of generic
branes,
\begin{align}
\nonumber \langle
z_0,y_0|(-1)^{F^c}\tilde{q}^{L_0^c}\tilde{z}^{N_0^c}|z'_0,y'_0\rangle
\!\!\ &=\ \!\!\frac{2i}{k} \int de' dn'
e^{i(n'-\frac12)(y'_0-y_0)+ie'(z'_0-z_0)}\sin(\pi e'/k)
   \hat{\chi}_{\langle e',n'\rangle }(\tilde{\mu},\tilde{\tau})\\[2mm]
\ &=\ \hat{\chi}_{\langle e,n\rangle }(\mu,\tau)\ -
                \hat{\chi}_{\langle e,n+1\rangle }(\mu,\tau)\,,
\label{eq:gg}
\end{align}
where the momenta $e,n$ are related to the coordinates of the
branes according to
$$ e\ =\ \frac{k(y'_0-y_0)}{2\pi} \ \ \ \ , \ \ \
n\ =\ \frac{k(z'_0-z_0)}{2\pi}-\frac{y'_0-y_0}{2\pi}\ .$$

Let us now turn to the boundary states of non-generic untwisted
branes in the \GL\ WZNW model. The boundary states of elementary bra\-nes associated with
non-generic position parameters $z_0$ and $y_0=2\pi s, s \in
\mathbb{Z},$ are given by
\begin{equation}
    \begin{split}
        |z_0;s\rangle & =\ \frac{1}{\sqrt{2ki}}
        \int_{e \neq mk} dedn\ \exp\bigl(2\pi i (n-1/2) s +
   iez_0\bigr)\  \sin^{-1/2}(\pi e/k)\
         |e,n\rrangle\  \\[2mm] & \hspace*{2cm} - \,
       \frac{1}{\sqrt{2\pi i}}
       \, \sum_{m\in\Z} \int dn\ \exp\bigl(2\pi i (n-1/2) s +
   imk z_0\bigr)\ |n\rrangle^{(m)}\ \ .
    \end{split}
\end{equation}
We also here verify that the proposed boundary states produce a consistent
open string spectrum by calculating the overlap between two non-generic
boundary states $|z_0;s\rangle$ and $|z_0';s'\rangle$,
\begin{equation} \label{eq:gng}
    \begin{split}
        \langle z_0;s|(-1)^{F^c}\tilde{q}^{L_0^c}\tilde{z}^{N_0^c}|z'_0;s'\rangle
\ &= \ \int  \frac{de'dn'}{2ki}\, \frac{e^{2\pi i(n'-1/2)(s'-s) +
ie'(z_0'-z_0)}}{\sin(\pi e'/k)} \ \hat{\chi}_{\langle
e',n'\rangle} (\tilde{\mu},\tilde{\tau})\\[2mm]
 &=  \ \hat{\chi}_{\langle n \rangle}^{(m)}(\mu,\tau)\,,\\
\end{split}
\end{equation}
where the labels $n$ and $m$ in the
character are related to the branes' parameters through
\begin{equation}
n \ = \ \frac{k(z'_0-z_0)}{2\pi} + s-s' \ \ \ \ , \ \ \ m = s'-s\ \ .
\end{equation}
The superscript on the character $\hat{\chi}_{\langle n
\rangle}^{(m)}(\mu,\tau)$ again means we have used the spectral flow
$\ga_m$ see ref.~\cite{Creutzig:2007jy} and appendix~\ref{app:representation}.
The following limit for $t$ any integer shows that in equation ~\eqref{eq:gng} is indeed a hidden $\tau$-dependence
\begin{equation}
    \lim_{e\rightarrow mk}\frac{1}{2ki}\int dn\ \frac{e^{2\pi itn}}{\sin(\pi e/k)} \ \hat{\chi}_{\langle
    e,n\rangle} (\tilde{\mu},\tilde{\tau}) \ = \ \int dn\ \tau\, e^{2\pi itn}\hat{\chi}_{\langle n\rangle}^{(m)} (\tilde{\mu},\tilde{\tau})\,.
\end{equation}
Thus we observe that the Ishibashi state $|n\rrangle$ \eqref{eq:ishiamplin} with its $\tau$-dependence is the natural atypical Ishibashi state contributing to the atypical boundary state.

Further, the overlap between a generic and a non-generic state is
\begin{equation}
    \begin{split}
       \langle z_0,y_0|(-1)^{F^c}\tilde{q}^{L_0^c}\tilde{z}^{N_0^c}|z'_0;s\rangle
\ &= \ \int  \frac{de'dn'}{k}\, e^{ i(n'-1/2)(2\pi s-y_0) + ie'(z_0'-z_0)} \ \hat{\chi}_{\langle
e',n'\rangle} (\tilde{\mu},\tilde{\tau})\\[2mm]
 &= \ \hat{\chi}_{\langle e,n \rangle}(\mu,\tau)\,, \\
\end{split}
\end{equation}
where
\begin{equation}
    n \ = \ \frac{k(z'_0-z_0)}{2\pi} + \frac{y_0}{2\pi}-s+\frac{1}{2} \ \ \ \ , \ \ \ \frac{e}{k}\ = \ s-\frac{y_0}{2\pi}\ \ .
\end{equation}

\subsection{Twisted boundary state}

The group of outer automorphisms of the Lie superalgebra \gl\ is of order 2. We already discussed the boundary states belonging to the trivial one.
The non-trivial one defines the following gluing conditions on the currents
\begin{equation}
    J^{E}\ = \ -\bar{J}^E \ \ \ , \ \ \ J^{N}\ = \ -\bar{J}^N \ \ \ , \ \ \
    J^{+}\ = \ -\bar{J}^- \ \ \ , \ \ \ J^{-}\ = \ \bar{J}^+\qquad{\rm for}\ z\ = \ \zbar\, .
\end{equation}
This translates into Neumann conditions for the bosonic and the fermionic fields, that is
\begin{equation}
    \del_n Y\ = \ \del_n Z \ = \ 0\qquad{\rm for}\ z\ = \ \zbar\,
\end{equation}
implying especially that the left movers of $Y$ coincide with its right movers up to the zero modes
\begin{equation}
    Y^L - Y^R \ = \ Y^L_0 - Y^R_0 \qquad{\rm for}\ z\ = \ \zbar\, .
\end{equation}
Thus the gluing conditions for the fermions are
\begin{equation}
    e^{Y_0^L}\del\chi^1\ = \ e^{Y_0^R}\delbar\chi^2\qquad,\qquad e^{-Y_0^L}\del\chi^2\ = \ -e^{-Y_0^R}\delbar\chi^1 \qquad{\rm for}\ z\ = \ \zbar\, .
\end{equation}

The boundary state $|\,\Omega\rrangle$ is easily constructed as before. It
has to satisfy
\begin{equation} \label{gluetwisted}
    \begin{split}
    (Y_n^L + Y^R_{-n})\, |\, \Omega\rrangle \ =(p_Y^L + p^R_{Y})\, |\, \Omega\rrangle \ =\ 0\,, \\
    (Z_n^L + Z^R_{-n})\, |\, \Omega\rrangle \ =(p_Z^L + p^R_{Z})\, |\, \Omega\rrangle \ =\ 0\,, \\
    (e^{Y_0^L}\chi^1_n + e^{Y_0^R}\bar{\chi}^2_{-n})\, |\, \Omega\rrangle \ =\ 0\,, \\
    (e^{-Y_0^L}\chi^2_n - e^{-Y_0^R}\bar{\chi}^1_{-n})\, |\, \Omega\rrangle \ =\ 0 \,,\\
\end{split}
\end{equation}
which can be computed to be
\begin{equation} \label{boundtwisted}
                |\, \Omega\,\rrangle \ = \ \sqrt{\pi/i}\exp \Bigl( \sum_{n=1}^\infty
\frac{1}{n} \bigl(Y^L_{-n}Z^R_{-n} + Z^L_{-n}Y^R_{-n}
- e^{Y^R_0-Y^L_0}\chi^2_{-n} \bar{\chi}^2_{-n} + e^{Y^L_0-Y^R_0}\chi^1_{-n} \bar{\chi}^1_{-n}\bigr)\Bigr)|0,0\,\rangle \, .
\end{equation}
Here, $|0,0\,\rangle$ denotes the vacuum defined by
$\chi_n^a|0,0\,\rangle=0$ for $n\geq0$ and
$Z^{L,R}_n|0,0\,\rangle=Y^{L,R}_n|0,0\,\rangle=p_Y^{L,R}|0,0\,\rangle=p_Z^{L,R}|0,0\,\rangle=0$
for $n>0$. The dual boundary state is constructed analogously.

Our main aim now is to compute some non-vanishing overlap of the twisted
boundary state $|\Omega\rrangle$. This requires the insertion of the
invariant bulk field $\chi^1\chi^2$, i.e.
\begin{equation}
    \begin{split}\label{eq:amplitude}
 \llangle\Omega\, |\, \tilde q^{L_0^c}(-1)^{F^c}\,
   \tilde z^{N_0^c}\, \chi^1\chi^2
         \, |\, \Omega\rrangle\, =\ \frac{\pi}{2k}\int de dn \ \frac{\hat{\chi}_{\langle e,n\rangle}
        (\tau,\mu)} {\sin(\pi e/k)} \ .
    \end{split}
\end{equation}
where $L_0^c = (L_0+\bar{L}_0)/2$ and $N_0^c = (N_0+\bar{N}_0)/2$ are
obtained from the zero modes of the Virasoro field and the current $N$.
Here the normalization in~\eqref{boundtwisted} by $\sqrt{\pi/i}$ was
important.\footnote{Note the difference of a $\sqrt{2}$ compared to
\cite{Creutzig:2008ek} which is due to a misprint there.}
This amplitude has been tested in detail in
\cite{Creutzig:2008ek}.

\subsection{Mixed amplitudes and their open strings}

Using the \GL-symplectic fermion correspondence we were able to construct
boundary states explicitly, and the amplitudes fits with the previously
known results calculated in \GL. The new explicit formulation also allows
us to compute new quantities such as overlaps for atypicals
\begin{equation}
    \begin{split}\label{eq:mixedatypicalamplitude}
  \llangle\Omega\, |\, \tilde q^{L_0^c}(-1)^{F^c}\,  \tilde z^{N_0^c}\,     \, |\, z_0;s\rrangle\  &=\
                  \sqrt{\frac{1}{2}}(-1)^s\, \prod\limits_{n=0}^\infty\, \frac{(1-\tilde q^n)}{(1+\tilde q^n)}\\
  &= \ (-1)^s\,q^{\frac{1}{32}}\,\prod\limits_{n=0}^\infty\, \frac{(1-q^{n+\frac{1}{4}}) (1-q^{n+\frac{3}{4}})}
          {(1-q^{n+\frac{1}{2}})^2}\,.\\
    \end{split}
\end{equation}
Note the independence on $z$, no matter whether we take $N_0^c$ as in the previous section or as in the untwisted case, which is natural since there does not exist a distinguished choice for $N_0^c$ for mixed amplitudes.

The corresponding open string theory is easily constructed using our
previous experience. That is, we demand untwisted gluing conditions on the
negative real line
\begin{equation}
    \begin{split}
        \del_u Y\ &= \ \del_u Z \ = \ 0\ \ , \\
        e^{Y_0^L}\del\chi^1 \ &= \ -e^{-Y_0^R}\bar{\del}\chi^1\ \ , \\
        e^{-Y_0^L}\del\chi^2 \ &= \ -e^{Y_0^R}\bar{\del}\chi^2  \qquad\qquad{\rm for}\ z\ = \ \bar{z} \qquad{\rm and}\ z+\bar{z}\ <\ 0\ ;\\
    \end{split}
\end{equation}
and twisted on the positive one
\begin{equation}
      \begin{split}
          \del_n Y\ &= \ \del_n Z \ = \ 0\ \ , \\
        e^{Y_0^L}\del\chi^1 \ &= \ e^{Y_0^R}\bar{\del}\chi^2\ \ , \\
        e^{-Y_0^L}\del\chi^2 \ &= \ -e^{-Y_0^R}\bar{\del}\chi^1 \qquad\qquad{\rm for}\ z\ = \ \bar{z} \qquad{\rm and}\ z+\bar{z}\ >\ 0\ ,\\
    \end{split}
\end{equation}
Then the fermions have a monodromy of order four around the origin
\begin{equation}
    \del\chi^1(ze^{2\pi i}) \ = \ i\,\del\chi^1(z)\qquad,\qquad \del\chi^2(ze^{2\pi i}) \ = \ -i\,\del\chi^2(z)\,,
\end{equation}
and the bosons a monodromy of order two
\begin{equation}
    \del Y(ze^{2\pi i}) \ = \ -\del Y\qquad,\qquad \del Z(ze^{2\pi i}) \ = \ -\del Z(z)\, .
\end{equation}
Thus the fermions have mode expansion
\begin{equation}
    \begin{split}
    \chi^1(z) \ &= \  \sum_{n\,\in\,\Z+\frac{3}{4}}\frac{1}{n}\chi^1_nz^{-n}\,,\\
    \chi^2(z) \ &= \  \sum_{n\,\in\,\Z+\frac{1}{4}}\frac{1}{n}\chi^2_nz^{-n}\,,\\
\end{split}
\end{equation}
and the bosons
\begin{equation}
    \begin{split}
        Y(z) \ &= \ \sum_{n\,\in\,\Z+\frac{1}{2}}\, \frac{1}{n}\,  Y_n\, z^{-n}\,,  \\
        Z(z) \ &= \ \sum_{n\,\in\,\Z+\frac{1}{2}}\, \frac{1}{n}\,  Z_n\, z^{-n}\,,  \\
    \end{split}
\end{equation}
We define the ground state to be bosonic if $s$ (the position parameter of
the non-generic brane) is even and fermionic if it is odd. The partition
function is then
\begin{equation}
    \tr(q^{L_0}(-1)^F) \ = \ (-1)^s\,q^{\frac{1}{32}}\,\prod\limits_{n=0}^\infty\, \frac{(1-q^{n+\frac{1}{4}}) (1-q^{n+\frac{3}{4}})}
          {(1-q^{n+\frac{1}{2}})^2}\, .
  \end{equation}
 The amplitude involving typical fields requires as usual zero mode insertions, i.e.
\begin{equation}
    \begin{split}\label{eq:mixedtypicalamplitude}
  \llangle\Omega\, |\, \tilde q^{L_0^c}(-1)^{F^c}\,
  \tilde z^{N_0^c}\, \chi^1\chi^2   \, |\, z_0,y_0\rrangle\  &=\  \frac{\sqrt{2}\pi}{k}e^{-iy_0/2}\frac{\prod\limits_{n=0}^\infty (1-\tilde q^n)}{\prod\limits_{n=0}^\infty (1+\tilde q^n)}\\
  &= \ \frac{2\pi}{k}e^{-iy_0/2}\, q^{\frac{1}{32}}\,\prod\limits_{n=0}^\infty\, \frac{(1-q^{n+\frac{1}{4}}) (1-q^{n+\frac{3}{4}})}
          {(1-q^{n+\frac{1}{2}})^2}\,,\\
    \end{split}
\end{equation}
and its open string spectrum can be constructed as in the symplectic fermion case.

In summary, we have been able to give a complete discussion of Cardy
boundary states in the \GL\ WZNW model. This was only possible due to the
new formulation in terms of symplectic fermions. As a result, we saw that
indeed also for the Lie supergroup \GL\ Cardy's condition holds, i.e. any
amplitude of two boundary states indeed describes an open string spectrum.

\section{Outlook}

In this note, we have established a correspondence between the
Wess-Zumino-Novikov-Witten model on the Lie supergroup \GL\ and free
scalars plus symplectic fermions. This correspondence introduces a new
efficient way to study the WZNW model. A natural question is whether there
exist generalizations of the procedure. \GL\ is special in the sense that
it is level-independent, i.e. rescaling the current $J^E$ by $\lambda^2$,
the fermionic currents $J^\pm$ by $\lambda$ and leaving $J^N$ invariant
simply changes the level $k$ by a factor of $\lambda^2$. Because of this
peculiarity we do not expect our procedure to extend in full generality,
but still we believe that for other supergroups at special levels such a
prescription also applies. A first attempt would be to look for free field
descriptions. This one can do immediately by taking the free Gross-Neveu
model of a dimension one half vector transforming in the adjoint
representation of the desired supergroup similar to what was done by
LeClair for \GL~\cite{LeClair:2007aj}.

For standard groups the procedure could also be useful. In the case of the
$H_3^+$ model one can use the procedure to arrive at a model of two free
scalars and the Liouville action. However, the vertex operators will take
a very complicated form.

A main motivation to study the correspondence was to serve as a toy model
for more sophisticated dualities. The guideline in our approach was to
rewrite the \GL\ currents in such a form that they are very symmetric as
the currents of a Gross-Neveu model are. We hope that this guiding
principle can serve as an important step in understanding the dualities
between the OSp(2N+2$|$2N) Gross-Neveu model and the principal chiral
model of the supersphere $S^{2N+1|2N}$~\cite{Candu:2008yw}.

Finally, we used the correspondence to construct the boundary states of
\GL\ and verify Cardy's condition, completing the series of investigations
\cite{Creutzig:2007jy,Creutzig:2008ek}. Especially, we got a picture of atypical Ishibashi states and
their contributions. As in other logarithmic conformal field theories
there exists more than just one Ishibashi state corresponding to each
atypical representation. Their overlaps might give $\tau$-dependent
contributions, but these Ishibashi states only contributed to atypical
boundary states. Based on the insights of this note and former work, we
should be able to investigate boundary states of other Lie supergroups
such as SU(1$|$2).

We remark, however, that there is still one open problem for branes in
\GL. There exist branes whose geometry is not a superconjugacy class and
which are rather special since their spectra are representations that are
indecomposable but reducible. Further they are peculiar since their dual
states are projected out \cite{Creutzig:2008ag}. These objects are not
understood and it would be interesting to study these in the light of the
\GL-symplectic fermion correspondence.

\subsection*{Acknowledgements}

We would like to express our gratitude to Volker Schomerus and David
Ridout for interesting discussion. PBR and TC would respectively like to
thank the Niels Bohr Institute, Copenhagen, and ETH, Z\"{u}rich, for
hospitality during part of this work.

\appendix

\section{\label{sc:Theta}Some formulas concerning theta functions}

   Let us recall some facts about the theta function in one variable,
   a good reference is Mumford's first book \cite{Mumford:1983}.
   $\theta(\mu,\tau)$ is the unique holomorphic function on
   $\mathbb{C}\times\mathbb{H}$, such that
\begin{equation}
   \begin{split}
     \theta(\mu+1,\tau)
     &\ =\ \theta(\mu,\tau),\\[2mm]
     \theta(\mu+\tau,\tau)
     &\ =\ e^{-\pi i\tau}e^{-2\pi i \mu}\theta(\mu,\tau),\\[2mm]
     \theta(\mu+\frac{1}{2},\tau+1)
     &\ =\ \theta(\mu,\tau),\\[2mm]
     \theta(\mu/\tau,-1/\tau)
     &\ =\ \sqrt{-i\tau}e^{\pi i\mu^2/\tau}\theta(\mu,\tau)\,,\\[2mm]
     \lim_{\text{Im}(\tau)\rightarrow \infty}\theta(\mu,\tau)
     &\ =\ 1\ \ .
   \end{split}
\end{equation}
   The theta functions has a simple expansion as an infinite product,
\begin{equation}
   \theta(\mu,\tau)
   \ =\ \prod_{m=0}^\infty\bigl(1-q^m\bigr)
        \prod_{n=0}^\infty\bigl(1+u^{-1}q^{n+1/2}\bigr)\bigl(1+uq^{n+1/2}\bigr)\
\ ,
\end{equation}
   where $q=e^{2\pi i\tau}$ and $u=e^{2\pi i \mu}$. The following variant is of concern to us
   \begin{equation}
   \theta\Bigl(\mu-\frac{1}{2}(\tau+1),\tau\Bigr)/\eta(\tau)
   \ =\
   (1-u)q^{\frac{-1}{24}}\prod_{n=1}^\infty\bigl(1-uq^n\bigr)\bigl(1-u^{-1}q^n\bigr)\
\ .
\end{equation}
   Its behavior under modular $S$ transformations which send the
   arguments of the theta function to $\tilde{\tau}=-1/\tau$ and
   $\tilde{\mu}=\mu/\tau$ can be deduced from the properties above. One
   simply finds
\begin{equation}
    \begin{split}
   \theta\Bigl(\mu-\frac{1}{2}(\tau+1),\tau\Bigr)/\eta(\tau)
   \ &=\ -ie^{\pi i\mu}q^{-\frac{1}{8}}\theta\Bigl(\tilde{\tau}(\frac{1}{2}-\mu)-\frac{1}{2},\tilde{\tau}\Bigr)/\eta(\tilde{\tau})
            \tilde{q}^{\frac{1}{2}(\mu-\frac{1}{2})^2}\\
        \ &=\-ie^{\pi i\mu} q^{-\frac{1}{8}}\tilde{q}^{\frac{1}{2}(\mu-\frac{1}{2})^2-\frac{1}{24}}\prod_{m=0}^\infty
        \bigl(1-\tilde{q}^{n+1-\mu}\bigr)\bigl(1-\tilde{q}^{n+\mu}\bigr)\,.\\
    \end{split}
\end{equation}

\section{Representation theory of \GL}\label{app:representation}

We recall some facts of the representation theory of \agl. A more detailed
discussion is given in the Appendix of \cite{Creutzig:2007jy}.

A useful tool for the investigation of the affine Lie superalgebra \agl\
and its representations are automorphisms that do not leave the horizontal
subalgebra invariant, the spectral flow automorphisms. The relevant one
for our purposes \cite{Creutzig:2007jy}, $\gamma_m$, leaves the modes
$N_n$
   invariant and acts on the remaining ones as
\begin{equation}
   \label{eq:SFE}
   \gamma_m(E_n)
   \ =\ E_n+km\delta_{n0} \ , \ \ \
\gamma_m(\Psi^\pm_n)\ =\ \Psi^\pm_{n\pm m}\, .
\end{equation}
These transformations induce a modification
   of the energy momentum tensor
   \begin{equation}
   \gamma_m(L_n)\ =\ L_n+mN_n\ \ .
\end{equation}
The characters of two representations $\rho$ and $\gamma_m(\rho)$ that are
related by spectral flow satisfy
\begin{equation}
   \chi_{\gamma_m(\rho)}(\mu,\tau)\ =\ \chi_\rho(\mu+m\tau,\tau)\, .
\end{equation}

Finally, we state the relevant characters, the typical one is
\begin{equation}
   \label{eq:CharTyp}
   \hat{\chi}_{\langle e,n\rangle}(\mu,\tau)
   \ = \ \hat{\chi}_{\overline{\langle e,n\rangle}}(\mu,\tau)
   \ = \ u^{n-1}q^{\frac{e}{2k}(2n-1+e/k)+1/8}
   \theta\Bigl(\mu-\frac{1}{2}(\tau+1),\tau\Bigr)\bigr/\eta(\tau)^3\,
\end{equation}
and the atypical one is following \cite{Rozansky:1992td}
\begin{equation}
   \label{eq:CharAtyp}
   \begin{split}
     \hat{\chi}_{\langle n\rangle}^{(m)}(\mu,\tau)
     &\ =\ \sum_{l=0}^\infty\hat{\chi}_{\langle mk,n+l+1\rangle}(\mu,\tau)\\[2mm]
     &\ =\ \frac{u^{n}}{1-uq^m}\,\frac{q^{\frac{m}{2}(2n+m+1)+1/8}
         \theta\Bigl(\mu-\frac{1}{2}(\tau+1),\tau\Bigr)}{\eta(\tau)^3}\, .
   \end{split}
\end{equation}



\bibliographystyle{utphys}
\bibliography{Bibgl11}

\end{document}